\documentclass[aps,pra,reprint,groupedaddress,amsmath,amssymb,superscriptaddress,twocolumns]{revtex4-2}
\usepackage{graphicx}
\usepackage{amsmath}
\usepackage{bm}
\usepackage{hyperref}
\usepackage{xcolor}
\usepackage{mathtools}
\usepackage{float}
\usepackage{soul}
\usepackage{verbatim}
\colorlet{color1}{black}
\newcommand\mycolor{color1}
\begin{document}
\title{Self phase-matched broadband amplification with a left-handed Josephson transmission line} 
\author{C. Kow}
\affiliation{Department of Physics and Applied Physics, University of Massachusetts, Lowell, MA 01854, USA}
\author{V. Podolskiy}
\affiliation{Department of Physics and Applied Physics, University of Massachusetts, Lowell, MA 01854, USA}
\author{A. Kamal}
\affiliation{Department of Physics and Applied Physics, University of Massachusetts, Lowell, MA 01854, USA}
%
%
%
\begin{abstract}
Josephson Traveling Wave Parametric Amplifiers (J-TWPAs) are promising platforms for realizing broadband quantum-limited amplification of microwave signals. However, substantial gain in such systems is attainable only when strict constraints on phase matching of the signal, idler, and pump waves are satisfied -- this is rendered particularly challenging in the presence of nonlinear effects, such as self- and cross-phase modulation, which scale with the intensity of propagating signals. In this work, we present a simple J-TWPA based on `left-handed' (negative-index) nonlinear Josephson metamaterial, \textcolor{\mycolor}{which has phase matching native to its design}, precluding the need for any complicated circuit or dispersion engineering. The resultant efficiency of four-wave mixing process can implement gains in excess of 20~dB over few GHz bandwidths with much shorter lines than previous implementations. Furthermore, the intrinsic phase matching considerably simplifies the J-TWPA design and operation compared to the previous implementations based on `right-handed' (positive index) Josephson metamaterials, making the proposed architecture particularly appealing for integration with large superconducting architectures. The left-handed JTL introduced here constitutes a new modality in distributed Josephson circuits, and forms a crucial piece of the unified framework that can be used to inform the optimal design and operation of broadband microwave amplifiers.
\end{abstract}
\maketitle
%
\section{Introduction}
%
Josephson Parametric Amplifiers (JPA) are a key element for high-fidelity microwave signal processing, \cite{Yurke1989,Yamamoto2008,Kamal2009,Bergeal2010v2} 
enabling applications ranging from qubit readout \cite{Stehlik2020}, real-time quantum feedback, quantum metrology \cite{Didier2015,Martin2020} to quantum sensing. Conventional JPA designs, based on Josephson junction(s) integrated in a resonant circuit, realize standing wave amplification at a fixed frequency; while ease of design of such amplifiers have made them a standard functionality in microwave measurements, such lumped-element designs are typically limited to relatively small instantaneous bandwidths and dynamic range (input signal powers for which amplification remains linear). 
There have been numerous proposals in recent years based on impedance engineering \cite{Mutus2014,Roy2015}, nonlinearity engineering \cite{Liu2017,Frattini2018} and coherent feedback via auxiliary modes \cite{Metelmann2014, Kamal2017} that can partially alleviate this issue; however, the ultimate amplification bandwidth, and concomitant saturation input powers, in such modified-JPA designs are still limited by the bare resonance linewidth of the signal mode.
\begin{figure*}[t!]
\begin{minipage}[c]{\textwidth}
\includegraphics[width=0.95\textwidth]{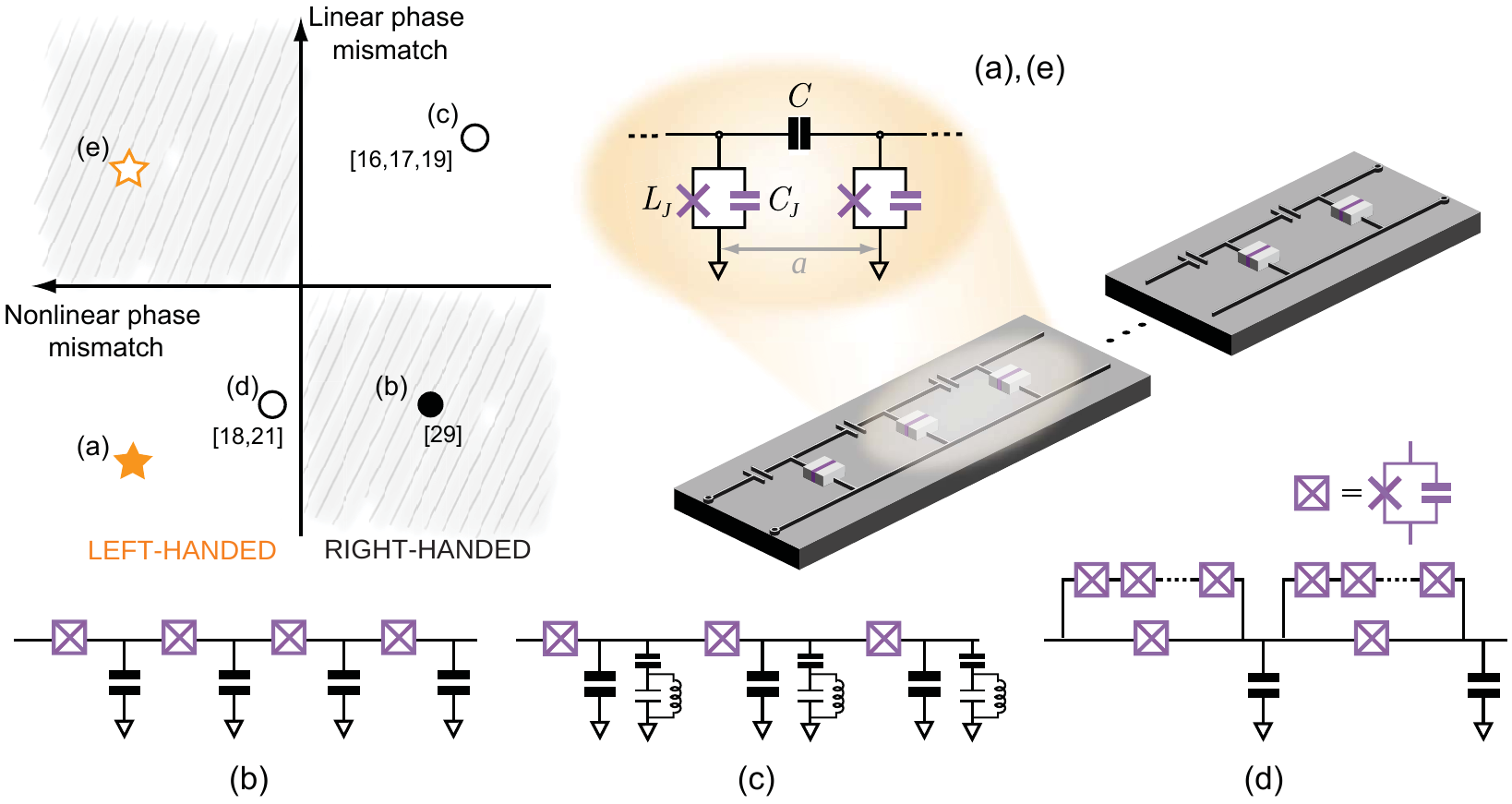}
\end{minipage}
\caption{Landscape of J-TWPA devices. The circles represent the design based on right-handed Josephson transmission lines while the stars denote the left-handed JTL designs proposed in this work. Filled symbols denote the `bare' right- and -left-handed J-TWPAs, while corresponding empty symbols indicate variations based on either linear or nonlinear phase engineering. The shaded quadrants show the regions where efficient amplification is not possible due to the linear and nonlinear dispersions having the same sign. (a) Schematic of a left-handed Josephson transmission line (JTL): each unit cell of length $a$ consists of a capacitance in series $C$, along with a Josephson junction acting as an inductive shunt to ground with junction inductance and capacitance denoted with $L_{J}$ and $C_{J}$ respectively. In the presence of a strong pump wave, a small probe signal injected at the input gets amplified as it travels down the JTL. Representative circuit designs of right-handed J-TWPAs are shown in panels (b), (c) and (d), along with respective references.}
\label{Fig:design}
\end{figure*}
\par
A compelling alternative to standing-wave JPAs are Josephson Traveling Wave Parametric Amplifiers (J-TWPA) [see \cite{Esposito2021} and references therein] that incorporate Josephson nonlinearity in a waveguide or transmission-line geometry. Unlike the lumped-element JPA circuits, the distributed nonlinearity of J-TWPA involves no resonating structures and thus, in principle, realizes much larger gain-bandwidth products. In addition to realizing broadband gain, TWPAs have the desirable property of unilateral amplification since only signals co-propagating (and hence phase matched) with the pump waves are amplified efficiently. This allows TWPAs to implement a natural separation of input and output channels, without involving any channel separation devices such as circulators or isolators that rely on external magnetic fields. Owing to this amenability and potential promise for integrated and scalable multiplexing, several theoretical and experimental approaches for realizing high-efficiency J-TWPAs have been explored, with the primary candidates being (i) engineered metamaterials based on arrays of Josephson junctions \cite{OBrien2014,White2015,Macklin2015,Bell2015,Miano2015,Grimsmo2021,Ranadive2021}, and (ii) nonlinear materials utilizing kinetic inductance of superconducting nanowires \cite{HoEom2012,Bockstiegel2014,Erickson2017,Adamyan2016,Vissers2016, Malnou2021}.
\par
Nonetheless, the very feature of phase matching that bestows TWPAs with their inherently non-reciprocal gain, also turns out to be the key challenge in the way of realizing gain over large propagation distances and device geometries. This is because the presence of strong Josephson nonlinearity makes the effective refractive index, and hence the phase difference between interacting signal and pump waves, intensity-dependent. Furthermore, since the signal amplitude scales with distance due to amplification, it becomes challenging to compensate for phase mismatch due to both linear and nonlinear dispersion between different propagating frequencies throughout the propagation distance \cite{AGRAWAL2019401,Yaakobi2013}. Inspired by ideas used in traveling-wave fiber-optic amplifiers, recent studies have explored solutions such as dispersion engineering of J-TWPAs \cite{OBrien2014,White2015,Planat2020} that is rooted in modifying linear dispersion of the bare (unpumped) line to compensate for the nonlinear phase mismatch in the presence of the pump. Variations on this theme based on impedance engineering \cite{Zhao2019} and nonlinearity engineering \cite{Bell2015,Ranadive2021}  also remain an area of active research. Figure~\ref{Fig:design} summarizes the different approaches adopted till date along with a sketch of representative device designs. As evident, almost all such approaches require increased complexity in design and fabrication of the circuit, loss of frequency tunability, and/or longer device lengths. Specifically, since dispersion-engineered designs typically involve additional circuit elements employing lossy dielectrics, the cumulative loss scales with length of the TWPA thus trading off one limitation for the other! Such concerns are especially pertinent when employing TWPAs as quantum-limited detectors and sources of squeezed radiation, applications where both pump tunability and low loss are imperative.
\par
In this work, we propose a novel and simple design based on left-handed Josephson metamaterial that can realize low-noise broadband amplification \emph{without} any need for complicated nonlinearity or dispersion engineering. The operation of a left-handed J-TWPA as a broadband amplifier is rooted in the compensation of linear dispersion-induced phase mismatch between signal(idler) and pump waves with nonlinearity-induced phase mismatch, an effect enabled by opposing directions of phase and group velocities in a left-handed transmission line. While left-handed transmission lines have been explored in linear optical applications, such as sub-wavelength focusing \cite{Grbic2004} and resonance cone formation \cite{Balmain2002,Chshelokova2012}, their potential as nonlinear media for wave mixing has remained largely unexplored. Specifically, left-handed Josephson transmission lines (JTLs) constitute a new modality in microwave superconducting circuits, where all designs explored to date employ right-handed transmission lines embedded with Josephson junctions. As depicted in Fig.~\ref{Fig:design}, left-handed JTLs effectively double the engineering landscape for J-TWPAs, introducing new operational regimes such as the `reversed dispersion' [(e) in Fig.~\ref{Fig:design}], which can be leveraged in combination of non-degenerate pumping for achieving a flat gain profile over a wide frequency range. 
%
%
%
\section{Equation of Motion for Left-Handed J-TWPA}
\label{Sec:Dandf}
%
%
\begin{figure*}[t!]
\centering
\includegraphics[width=0.9\textwidth]{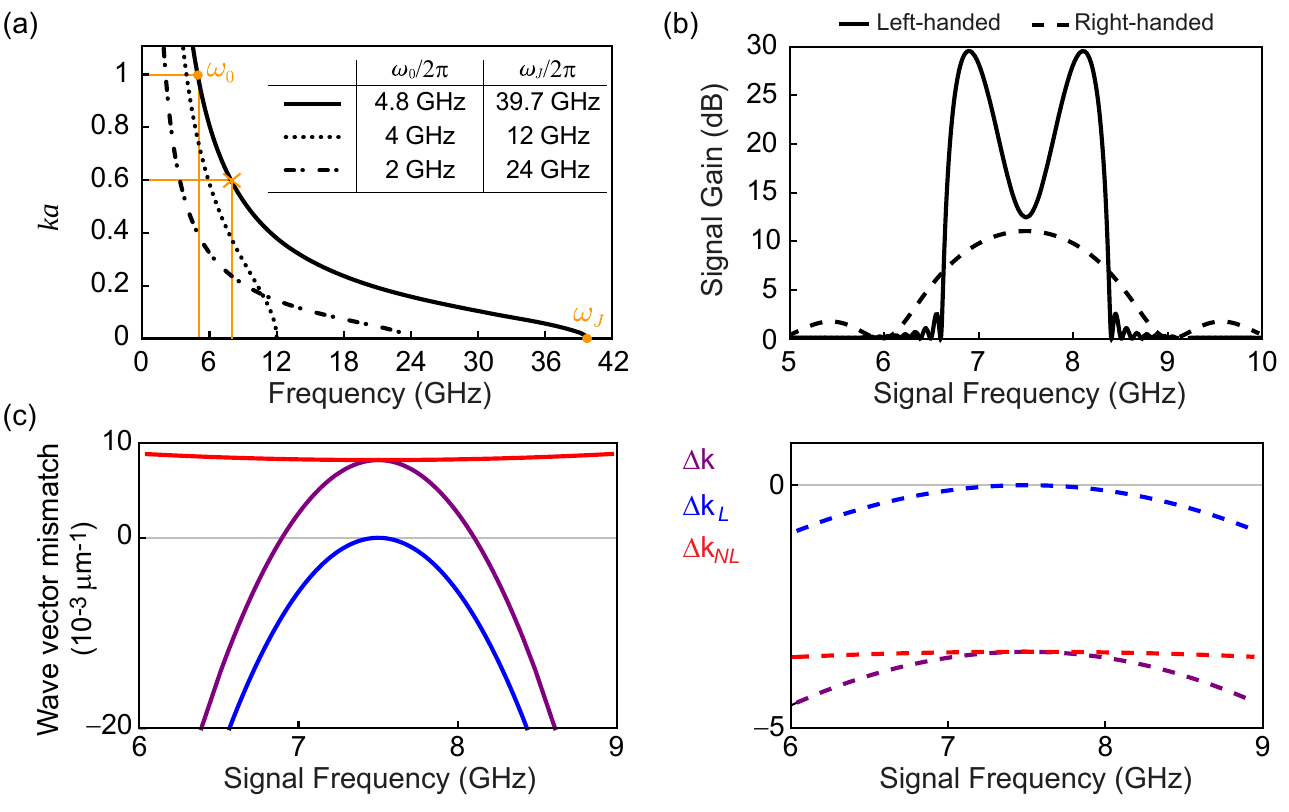}
 \caption{(a)~Dispersion relation for a left-handed JTL shown for three different choices of ($\omega_{J}$, \,$\omega_0$). The position of the cross denotes the choice of pump frequency, corresponding to wave vector $ka=0.6$, used for the calculations presented in the following panels.
 (b)~Comparison of frequency-dependent gain between left-handed ($l=1000a$) and right-handed ($l=2000a$) J-TWPAs [Figs.~\ref{Fig:design}(a)-(b)]. The circuit parameters used for the left-handed J-TWPA are $L_{J} = 1670~\text{pH}$, $C_{J} = 9.6~\text{fF}$, $C = 667~\text{fF}$, while those for the right-handed J-TWPA are $L_{J} = 100~\text{pH}$, $C_{J} = 329~\text{fF}$, $C = 39~\text{fF}$. For both calculations, $\omega_p/2\pi = 7.5~\text{GHz}$, $a = 10~\mu\text{m}$, $I_p = 0.5I_0$, and $Z_{\text{c}} = \sqrt{L_{J}/C} = 50~\Omega$ were used. The peak gain of 30 dB (11 dB) is realized at relative detuning of $\delta^{\rm max} = 0.6$ ($\delta^{\rm max} = 0$) in the left-handed (right-handed) design.
 (c) Comparison of frequency-dependent wave vector mismatch between the left-handed (left) and right-handed (right) J-TWPAs. 
 }
 \label{Fig:fig2}%
\end{figure*}
As shown in Fig.~\ref{Fig:design} the primary difference, between a left-handed Josephson transmission line (JTL) and the previous TWPAs designs designs based on right-handed JTL, is the exchange of the inductive ($L_{J}$) and capacitive ($C$) elements to be the parallel and series impedance elements in the transmission line respectively. The additional shunt capacitance $C_{J}$ denotes the intrinsic Josephson capacitance. The corresponding linear dispersion relation for the line is given by [Fig~\ref{Fig:fig2}(a)],
\begin{align}
    k_{m} &= \frac{\omega_{0}\sqrt{1-\omega_{m}^{2}/\omega_{J}^{2}}}{a\omega_{m}},
    \label{Eq:disp}
\end{align}
where $a$ denotes the size of a unit cell, $\omega_m$ is the frequency of the propagating wave, $\omega_J = 1/\sqrt{L_{J}C_{J}}$ is the Josephson plasma frequency that sets the cut-off frequency of the JTL waveguide, and $\omega_{0} = 1/\sqrt{L_{J}C}$ is frequency corresponding to $k_{m}a=1$. It is worthwhile to note that in a left-handed JTL the directions of wave and energy propagation are anti-parallel, given the sign difference between the wave and group velocities,
\begin{subequations}
\begin{align}
    v_w(\omega_{m})  &=\frac{a\omega_{m}^{2}}{\omega_{0}\sqrt{1-\omega_{m}^{2}/\omega_{J}^{2}}} \approx \frac{a\omega_{m}^{2}}{\omega_{0}} > 0, \label{Eq:pvel} \\
    v_g(\omega_{m})
    &=-\frac{a\omega_{m}^{2}\sqrt{1-\omega_{m}^{2}/\omega_{J}^{2}}}{\omega_{0}} \approx -v_w(\omega_{m}) < 0.
    \label{Eq:gvel}
\end{align}
\label{Eq:2}%
\end{subequations}
%
%
%
\subsection{Linear Amplification}
%
\textcolor{\mycolor}{In order to derive the wave equation for a left-handed JTL, we expand the Josephson cosine potential and retain only the leading order nonlinear term.} Such a `perturbative' approach is reasonable to describe the regime of linear amplification, when the current flowing through the junction remains small as compared to the critical current of the junction ($I_{0}$). This leads to the following equation of motion for the for the propagating field amplitude, described in terms of the position-dependent flux variable $\phi (x,t)$ (Supplementary Section I.A),
\begin{align}
    C_{J}\frac{\partial^{2}\phi (x,t)}{\partial t^{2}} - Ca^{2}\frac{\partial^{4}\phi(x,t)}{\partial x^{2}\partial t^{2}} + \frac{\phi(x,t)}{L_{J}} - \frac{\phi^{3}(x,t)}{6I_{0}^{2}L_{J}^{3}} = 0.
    \label{Eq:motion}
\end{align} 
The first three terms in Eq.~(\ref{Eq:motion}) describe the linear propagation, while the fourth term describes the nonlinear frequency mixing. It is worth noting that the nonlinearity appears directly in the flux amplitude (i.e. potential energy) here, unlike the right-handed J-TWPA where the nonlinearity appears in the kinetic energy term as $(\partial^{2}\phi/\partial x^{2}) (\partial \phi/\partial x)^{2}$ \cite{Yaakobi2013,OBrien2014}.
In order to derive linear amplification response of the left-handed JTL, we then develop the solution of the form 
\begin{equation}
   \phi(x,t) = \sum_{m \in \{p,s,i\}} \frac{A_m (x)}{2}  e^{i(k_{m}x-\omega_{m}t)} + c.c.
    \label{Eq:sol}
\end{equation}
where $m \in \{p,s,i\}$ indexes the pump, signal and idler waves respectively. 
%
%
Note that in contrast to previous theoretical proposals considering nonlinear optics of bulk metamaterials \cite{Popov2006,Litchinitser2007}, all three waves supported by the left-handed J-TWPA have negative group velocity. Performing harmonic balance dictated by energy conservation in the four-wave mixing process ($2\omega_p = \omega_s + \omega_i$), and solving the resultant coupled system of equations leads to the following expression for signal/idler amplitudes (Supplementary Section I.B), 
\begin{eqnarray}
    \tilde{A}_{s,i} (x) &= \left\{\sqrt{G_{c}}e^{-i\phi}\tilde{A}_{s,i}(0) + \sqrt{G_{t}}\tilde{A}_{i,s}(0)\right\}e^{i\Delta k/2}.
    \label{Eq:amplitude}
\end{eqnarray}
Here $G_{c}$ and $G_{t}$ represent the $\emph{cis}$- and $\emph{trans}$-gain of the amplifier respectively,
\begin{subequations}
\begin{align}
    &  G_{c}(x) = 1 + \sinh^{2}(gx)\left(1 + \left(\frac{\Delta k}{2g}\right)^{2}\right), \label{Eq:sgain} \\
    &  G_{t}(x) =(\beta_{s,i}/g)^{2}\sinh^{2}(gx), \label{Eq:transgain} \\
    &  \phi = \text{tan}^{-1}\left(\frac{\Delta k}{2g}\text{tanh}(gx)\right), \label{Eq:phi}
\end{align}
\label{Eq:cistransgain}%
\end{subequations}
with $g=\sqrt{\beta_s\beta_i^{\star} - \left(\Delta k/2\right)^2}$ denoting the gain per unit length of the amplifier, $\beta_{s,i}$ being the nonlinear coupling per unit length for signal and idler waves, and $\Delta k$ being the total wave vector mismatch between the wave propagating at pump frequency and a fixed signal (idler) frequency. 
%
%
%
%
%
%
\par
Figure~\ref{Fig:fig2}(b) contrasts $G_{c}(x)$ calculated for left-handed and right-handed J-TWPAs comprising $1000$ and $2000$ unit cells respectively, both pumped at a frequency $\omega_p/2\pi = 7.5~\text{GHz}$. The stark difference in both the magnitude and profile of the frequency-dependent gain for left- vs right-handed designs is rooted in the frequency profiles of respective $\Delta k(\omega)$, which gets  contributions both from the linear dispersion of the JTL $\Delta k_L(\omega)$, and the nonlinearity-induced self- and cross-phase modulation of the propagating signals $\Delta k_{NL}(\omega)$, i.e.
$
    \Delta k(\omega) = \Delta k_L(\omega) + \Delta k_{NL}(\omega).
$
Notably, these two contributions are of opposite signs in a left-handed J-TWPA as shown in Fig.~\ref{Fig:fig2}(c), since the signs of the $\Delta k_L(\omega)$ and $\Delta k_{NL}(\omega)$ are determined by the wave velocity $v_{w}(\omega)$ and group velocity $v_{g}(\omega)$ respectively, 
\begin{subequations}
\begin{align}
\Delta k_{L}(\delta)a &= \left(\frac{2\omega_{p}}{v_{w}(\omega_{p})} - \frac{\omega_{s}}{v_{w}(\omega_{s})} - \frac{\omega_{i}}{v_{w}(\omega_{i})}\right)a\nonumber\\
& \approx - \frac{2\omega_{0}}{\omega_{p}} \left(\frac{\delta^{2}}{1 - \delta^{2}}\right) < 0,\\
\Delta k_{NL}(\delta) a &= 2\rho \left(\frac{\omega_{p}}{v_{g}(\omega_{p})} - \frac{\omega_{s}}{v_{g}(\omega_{s})} - \frac{\omega_{i}}{v_{g}(\omega_{i})}\right)\nonumber\\
 &\approx 2\rho  \frac{\omega_{0}}{\omega_{p}}\left(\frac{1 +\delta^{2}}{1 - \delta^{2}}\right) >0,
\end{align}
\label{Eq:knlkl}%
\end{subequations}
where, in the second step of each equation, we have used Eqs.~(\ref{Eq:2}) and parametrized signal/idler frequencies in terms of a dimensionless detuning $\delta$ from the pump, ${\omega_{s,i} =\omega_{p} (1 \pm \delta)},\,-1<\delta < 1$. Here, ${\rho= \left(I_p/I_0\right)^{2}\left(\omega_{0}/4\omega_{p}\right)^{2}}$ is the nonlinear mixing coefficient determined by the pump amplitude and frequency (Supplementary Section I.B). As evident from Eqs.~(\ref{Eq:knlkl}), the relative sign difference between wave and group velocity in a left-handed metamaterial enables a left-handed J-TWPA to be phase-matched over a broad bandwidth \footnote{Higher-order nonlinear effects such as group velocity dispersion need to be accounted for while analyzing high-pump power regime.}. 
\par
In contrast, in a right-handed J-TWPA both $v_{w}$ and $v_{g}$, and hence linear and nonlinear wave vector mismatch, are restricted to be of the same sign owing to the always-convex dispersion of a right-handed JTL. To circumvent this issue, several approaches centered on either modifying $\Delta k_{L}$\cite{OBrien2014} or $\Delta k_{NL}$ \cite{Bell2015,Ranadive2021} have been explored in designs based on right-handed JTLs. Besides employing complicated design engineering, such solutions necessarily lead to other constraints. For instance, a common and widely adopted approach involves compensating the mismatch by periodically loading the line with resonant elements in order to open a band gap in $\Delta k_{L}$ [Fig.~\ref{Fig:design}(c)]; this, however, leads to limited frequency tunability of the J-TWPA since now the pump needs to be tuned to be near the dispersion feature to avail of the intended linear mismatch to be compensated by the nonlinear mismatch in the presence of the pump. Similarly, engineering $\Delta k_{NL}$ via SQUID-based designs require additional lines for flux control and complicated pump engineering. In contrast, \textcolor{\mycolor}{the native phase matching property} of left-handed J-TWPA significantly simplifies its design and operation. Furthermore, the resultant efficiency of the four-wave mixing process leads to a peak gain that scales exponentially with length of the line [unlike quadratic scaling in the `bare' right-handed JTL, Fig.~\ref{Fig:design}(b)], allowing usage of shorter lines which translates to reduced distributed loss and tighter fabrication control. Such considerations become especially crucial while evaluating prospects of J-TWPAs as sources of broadband squeezed radiation, where frequency-dependent line loss can severely limit the achievable squeezing \cite{Grimsmo2017,Houde2019}. 
%
%
%
\subsection{Unique features of Left-handed J-TWPA}
\label{Sec:unique}
%
In addition to phase matching over broad bandwidths, there are several unique features of left-handed J-TWPAs which we discuss in the following sections.
%
\subsubsection*{Peak Gain and Dynamic Range}
\label{Sec:gainsat}
%
\begin{figure}[t!]
\centering
\includegraphics[width=0.45\textwidth]{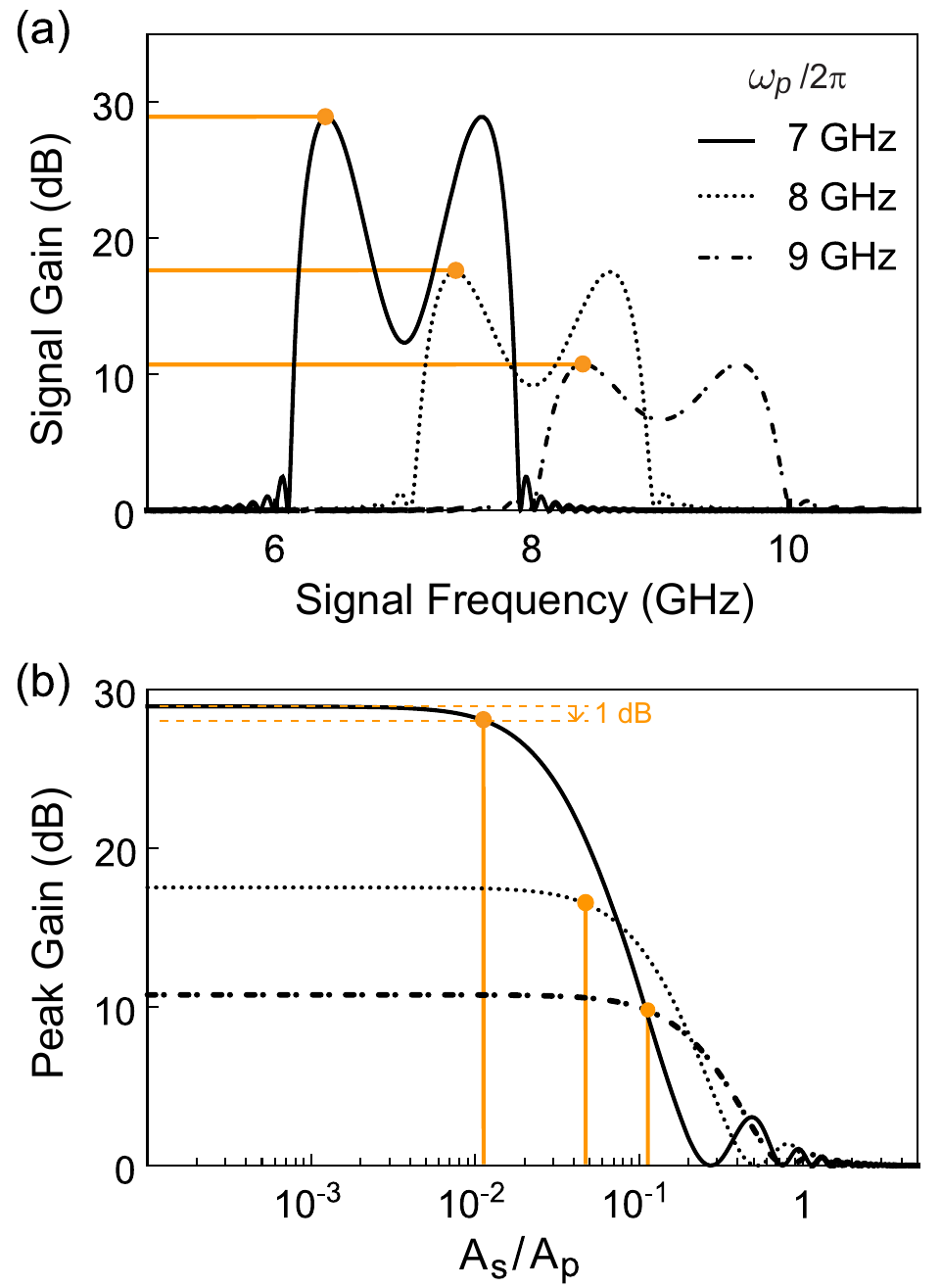}
 \caption{(a) Frequency-dependent $G_{c}$, calculated for a left-handed J-TWPA with 800 unit cells, for three different choices of pump frequencies. The rest of the circuit parameters are same as those used in Fig.~\ref{Fig:fig2}(a). Peak gains of $29\,\text{dB}$, $17.5\,\text{dB}$ and $10.7\,\text{dB}$, estimated for $\omega_p/2\pi = 7\,\text{GHz}$, $8\,\text{GHz}$, and $9\,\text{GHz}$ respectively, are highlighted with horizontal orange lines. (b)~Variation of peak gain with normalized input signal amplitude $A_{s}/A_p$. The orange dots indicate $1\,\text{dB}$ gain compression on each curve; the x-intercepts indicated with vertical orange lines, ${A_s/A_p = 0.011, \,0.046, \,0.12}$, represent the dynamic ranges for the three pump frequencies indicated in (a).}
 \label{Fig:fig3}
\end{figure}
The peak gain of 30 dB in Fig.~\ref{Fig:fig2}(b) is realized at a frequency corresponding to a perfect wave matching condition, $\Delta k(\delta) = \Delta k_{L}(\delta) + \Delta k_{NL}(\delta) = 0$. In view of Eqs.~(\ref{Eq:knlkl}), this corresponds to a relative detuning from the center (pump) frequency,
\begin{align}
    \delta^{\rm max} &\approx \left(\frac{\rho}{1 - \rho}\right)^{1/2}.
    \label{Eq:det}
\end{align}
Note that, unlike right-handed design, doubly degenerate condition, i.e. $\delta =0$, does not correspond to perfect phase matching in a left-handed J-TWPA. Furthermore, in the limit ${\rho \ll 1}$, the gain per unit cell at $\delta^{\rm max}$ simplifies to
$
    g a
\approx \rho \omega_0/\omega_{p} = \left(I_p/4I_0\right)^{2}\left(\omega_{0}/\omega_{p}\right)^{3} .
$
As depicted in Fig.~\ref{Fig:fig3}(a), this leads to a sharp increase in peak gain with decrease in pump frequency. \textcolor{\mycolor}{This is another contrasting feature from right-handed J-TWPA, where higher gain is realized for higher pump frequencies \cite{OBrien2014,Yaakobi2013, Ranadive2021, Bell2015}. A critical advantage of this low-pass filtering property of the left-handed Josephson transmission line is the suppression of higher harmonic generation (Supplementary Section 4.A). Significant engineering efforts are being expended to eliminate the intermodulation products and sidebands mediated by these higher pump harmonics, that are known to decrease quantum efficiency and effective squeezing attainable with right-handed J-TWPAs \cite{Peng2021}.}
\par
The analysis until now assumes a ``stiff" or undepleted pump amplitude irrespective of the signal gain, which strictly holds true under small signal approximation. However, as signal amplitude increases, either at the input or due to amplification down the line, and becomes comparable to pump amplitude, the amplifier response becomes nonlinear and pump depletion effects due to signal (idler) backaction need to be considered \cite{Kamal2009}. Figure~\ref{Fig:fig3}(b) plots the result of a calculation including these effects, showing that gain of a left-handed J-TWPA rolls-off at high signal powers. Further, the gain compression sets in earlier when the same device is operated at lower pump frequencies. This is in accordance with the standard amplifier physics that higher gain leads to saturation for smaller signal powers \cite{Abdo2013}, in conjunction with the fact that a left-handed J-TWPA realizes higher gain at lower $\omega_{p}$. \textcolor{\mycolor}{For a detailed comparison of dynamic range between the two J-TWPA modalities, we refer the reader to Supplementary Section 2.B.}
%
%
\subsubsection*{Zero-dispersion frequency}
\label{Sec:ndgpumping}
%
\begin{figure}[t!]
\centering
\includegraphics[width=0.5\textwidth]{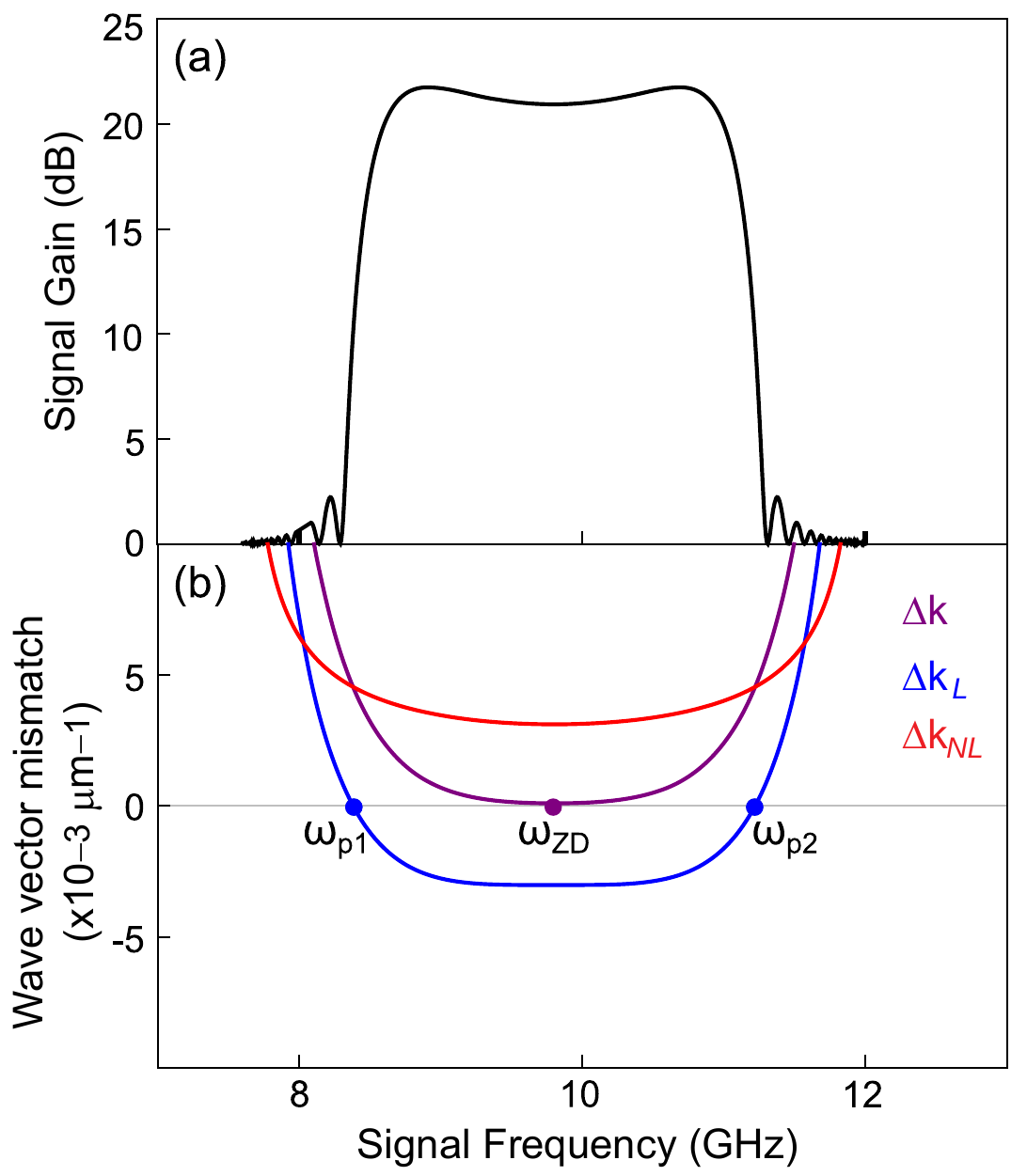}
\caption{(a) Frequency-dependent gain of a left-handed J-TWPA with 1650 unit cells, double-pumped near ${\omega_{\rm ZD}/2\pi = 9.8\,\text{GHz}}$ with two non-degenerate pumps detuned by $\pm 1.42\,\text{GHz}$. The circuit parameters correspond to the dotted dispersion curve in Fig.~\ref{Fig:fig2}(a), $L_{J} = 1989.4~\text{pH}$, $C_{J} = 88.4~\text{fF}$, $C = 795.8~\text{fF}$, $I_p = 0.5I_0$, and $Z_{\text{c}} = 50~\Omega$. (b) Frequency-dependent phase mismatch showing $\Delta k_L = 0$ at the two pump frequencies $\omega_{p1}/2\pi = 8.38\,\text{GHz}$ and $\omega_{p2}/2\pi = 11.22\,\text{GHz}$.}
 \label{Fig:4}
\end{figure}
Another feature unique to the left-handed J-TWPA is the existence of a ``zero-dispersion frequency", $\omega_{\rm ZD}$. 
%
%
The location of the $\omega_{\rm ZD}$ corresponds to an inflection point of the dispersion relation, which leads to the following useful expression for its location,
$
    \omega_{\rm ZD} = \sqrt{2/3} \omega_{J}.
$
On the other hand a right-handed JTL, given its always-convex dispersion, cannot support a $\omega_{\rm ZD}$.
\begin{figure*}[t!]
\centering
\includegraphics[width=0.95\textwidth]{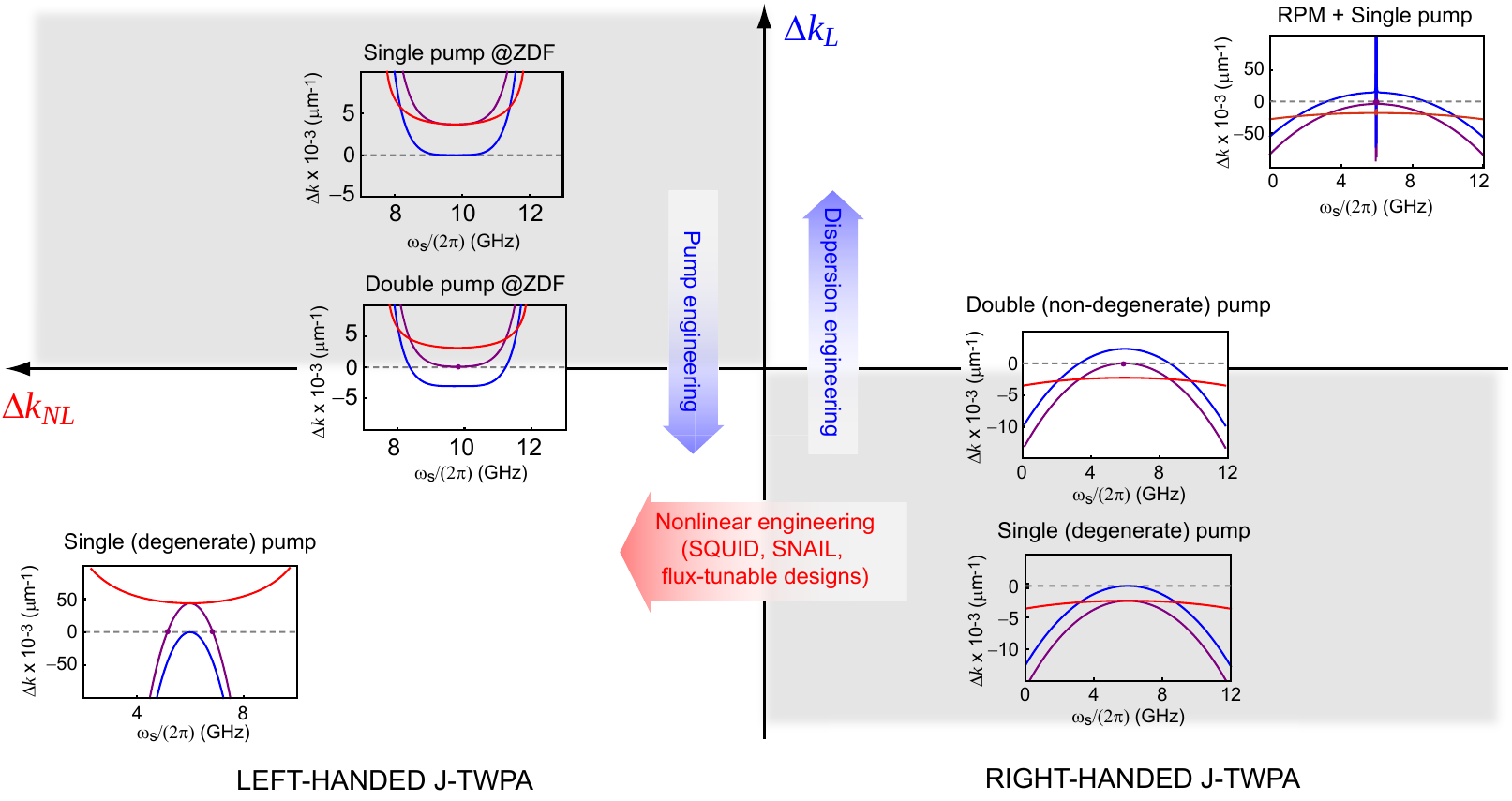}
 \caption{Classification of J-TWPA designs in $(\Delta k_{NL}, \Delta k_L)$ plane. Different J-TWPA designs are indicated with the representative phase mismatch plots, showing linear (blue), nonlinear (red) and total (purple) wave vector mismatch. The gray quadrants represent the regions where $\text{sgn}(\Delta k_L)=\text{sgn}(\Delta k_{NL})$ making it impossible to achieve perfect phase matching ($\Delta k = 0$). The labeled blue and red arrows show the movement in this plane based on the nature of engineering solution.}
 \label{Fig:fig5}
\end{figure*}
\par
Further, $\Delta k_L$ flips sign as pump frequency is scanned across $\omega_{\rm ZD}$ in a left-handed J-TWPA. Since $\Delta k_{NL} > 0$ in a left-handed J-TWPA, phase matching for signal frequencies in the range $\omega_{\rm ZD} \leq \omega < \omega_{J}$ seems a nonstarter (Supplementary Section III)). In fact, this regime can be thought of as a perfect dual of the right-handed JTL where both $\Delta k_{L}$ and $\Delta k_{NL}$ were negative. Nonetheless, it is worthwhile to note that the linear dispersion also has the flattest profile near $\omega_{\rm ZD}$ since the curvature of the dispersion curves goes to zero at $\omega_{\rm ZD}$; this bodes well for achieving a flat broadband gain if phase matching could indeed be achieved. We show that the latter problem can be circumvented via the use of non-degenerate dual pumps \cite{Kamal2009} in Figure~\ref{Fig:4}(a) (Supplementary Section II). 
\par
As evident, a double-pumped left-handed J-TWPA achieves an almost flat gain profile, realizing a gain in excess of $20$~dB gain over a bandwidth of about $1.5~\text{GHz}$. The flat profile of the gain achieved with non-degenerate pumping is consistent with the shape of the corresponding phase mismatch curves shown in Fig.~\ref{Fig:4}(b). Note that double pumping nulls the linear dispersion, $\Delta k_L$, at the two pump frequencies leading to a $\Delta k_L < 0$ for $\omega_{p1} < \omega <\omega_{p2}$. Combined with a $\Delta k_{NL} > 0$, this leads to a perfect wave vector match at the center frequency $\omega_{\rm ZD} = (\omega_{p1} + \omega_{p2})/2 = 9.8\,\text{GHz}$ resulting in a flat broadband gain. \textcolor{\mycolor}{We note that the dynamic range in the two-pump non-degenerate pumping case remains the same as the single pump case for nominally identical small signal gain values.}
%
%
\section{Discussion}
\label{Sec:conclusions}
%
\textcolor{\mycolor}{In summary, we have presented a natively phase-matched left-handed J-TWPA platform which achieves low noise broadband amplification by exploiting the opposing signs of group and phase velocity in a left-handed Josephson transmission line (JTL).} The simplicity of the proposed design precludes the need for any complicated circuit or nonlinearity engineering, significantly easing the fabrication of J-TWPA devices.  In addition, it supports unique features such as existence of a reversed dispersion regime that can realize flat broadband gain by simply changing the operation frequency and employing non-degenerate pumps. 
\par
The principles developed in our work considerably expand the design landscape of generic traveling-wave amplifiers. Further, they provide a framework to inform design vs operational trade-offs, such as maximum gain at specific frequencies vs constant gain over wide bandwidths, in a given (right-handed vs left-handed) metamaterial platform. We elucidate this in Fig.~\ref{Fig:fig5} that incorporates several insights generated during the course of this research to present a unified view of several J-TWPA designs, each of which can be placed in the relevant quadrant depending on the relative sign difference between linear ($\Delta k_L$) and nonlinear ($\Delta k_{NL}$) wave vector mismatch. The aim of J-TWPA engineering is to be in the white quadrants of Fig.~\ref{Fig:fig5}, where perfect phase matching can be realized via optimization of design and/or operation of a given device. The `bare' left-handed J-TWPA already lies in a favorable quadrant in this plane, unlike the `bare' right-handed J-TWPA; a potential means to move between these quadrants is to modify the nonlinear phase contribution by means of engineering the Josephson nonlinearity, such as those employed in Refs.~\cite{Bell2015} and \cite{Ranadive2021}, though at the cost of more involved designs that employ flux-tunable Josephson circuits [Fig.~\ref{Fig:design}(d)]. 
\par
Similarly, linear phase engineering determines the movement along the y-axis, typically accomplished via either circuit engineering to modify the line dispersion \cite{OBrien2014} or pump engineering, as discussed in the case of a left-handed J-TWPA operated near $\omega_{\rm ZD}$. This representation also shows how naively combining two approaches can be detrimental to the cause: for instance, while double-pumping a bare right-handed J-TWPA improves phase matching, double-pumping a resonantly phase-matched (RPM) version can spoil it. This is because the advantage of engineering a positive $\Delta k_{L}$, via an RPM-induced band gap in the line dispersion, is negated by pinning the dispersion curve to be zero near the two pump frequencies. By simple inspection, one can see that given the convex dispersion for both the right-handed JTL, and left-handed JTL below $\omega_{\rm ZD}$, the $\Delta k_{L}$ is always pushed along the positive axis, while the opposite is true for a left-handed J-TWPA pumped above $\omega_{\rm ZD}$. Thus, the diagram serves as a useful guide in understanding the landscape of distributed amplifier designs, with the arrows indicating optimal strategies to achieve efficient broadband amplification through a combination of nonlinear (x-axis) and linear (y-axis) phase engineering.
\par
\textcolor{\mycolor}{Interestingly, even in terms of parameter optimization, left-handed J-TWPA perfectly complements its right-handed counterparts. Designing left-handed JTLs that operate in the standard 4-12 GHz frequency range of interest for applications such as qubit readout, while maintaining short line lengths (with less than 1000 unit cells) and being impedance matched to 50$\Omega$ external microwave circuitry, constrains the junction critical currents in the range of few hundred nAs; this makes the junction fabrication amenable to planar electron beam lithography instead of complicated trilayer fabrication needed for large junctions employed in right-handed designs. The price to pay is the reduced dynamic range which can be enhanced either by implementing the requisite inductance to using an array of larger junctions or switching to longer line lengths. Correspondingly, the inline capacitance is accordingly higher than the right-handed designs, which leads to higher overall loss in the left-handed designs; nonetheless, this again can be compensated by exploiting the higher efficiency of the natively phase-matched four-wave mixing process and using shorter lines (Supplementary Section 4.B). One other crucial advantage of the higher gain per unit length in the left-handed design is the enhanced robustness to disorder in the JTL \cite{Karlsson1998}.}
\par
Given the rapidly growing relevance of Josephson parametric circuits in platforms as diverse as superconducting qubits \cite{Aumentado2020}, semiconductor quantum dots \cite{Schaal2020}, quantum acoustics \cite{Andersson2021} and quantum optomechanics \cite{Heikkila2014}, the left-handed JTL investigated here provides new ground for exploration and optimization of such devices. Since linear left-handed transmission lines have recently been experimentally demonstrated in superconducting circuit platforms \cite{Wang2019}, the left-handed J-TWPA proposed here is within easy reach of current state-of-the-art in the field. In combination with upcoming ideas such as Floquet-mode pumping \cite{Peng2021} and designs employing novel nonlinear elements, such as SNAIL \cite{Sivak2019} and superinductances \cite{Masluk2012,Sivak2020}, we hope that the present work will help accelerate development of amplifiers that can address long-standing challenge of broadband squeezing, a critical functionality for both information processing and quantum sensing.
%
%
%
\begin{acknowledgments}
The authors wish to thank Tristan Brown and Jamie Kerman for comments on the manuscript, and especially Kevin O'Brien for insightful discussions and generously sharing the details of the open source JTWPA simulator. This research was supported by NSF under grant number DMR-2047357. V.P. acknowledges support from NSF under grant number DMR- 2004298.
\end{acknowledgments}
%
%
%
%
%
%
%
%
\end{document}


\title{Supplemental Material for\\
Self phase-matched broadband amplification with a left-handed Josephson transmission line} 
\author{C. Kow}
\affiliation{Department of Physics and Applied Physics, University of Massachusetts, Lowell, MA 01854, USA}
\author{V. Podolskiy}
\affiliation{Department of Physics and Applied Physics, University of Massachusetts, Lowell, MA 01854, USA}
\author{A. Kamal}
\affiliation{Department of Physics and Applied Physics, University of Massachusetts, Lowell, MA 01854, USA}
\maketitle
%
\section{Equations of motion for left-handed J-TWPA}
\label{App:waveeqn}
\subsection{Wave equation}
%
\begin{figure}[b!]
\centering
\includegraphics[width=0.9\columnwidth]{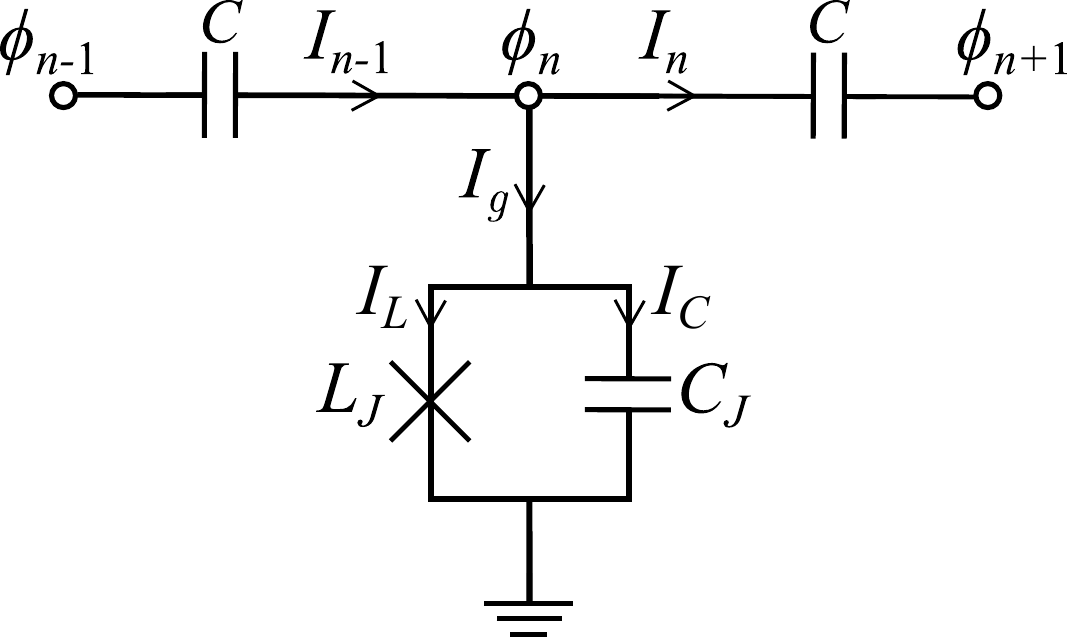}
 \caption{Circuit diagram of a single unit cell of a left-handed Josephson transmission line (JTL).}
 \label{Fig:A1}
\end{figure}
In this section, we present the derivation of nonlinear equation of motion describing field propagation in a left-handed JTL using the method of nodes. Using the Kirchoff's current conservation law at the $n^{\rm th}$ node for the circuit shown in Fig.~\ref{Fig:A1}, 
\begin{align}
    I_{n-1} &= I_{n} + I_{g} = I_{n} + I_{L} + I_{C}. \label{Eq:A1}
\end{align}
where
\begin{subequations}
\begin{align}
    & I_{n} = C\frac{d^{2}}{dt^{2}}\left(\phi_{n} - \phi_{n+1}\right), \label{Eq:A2a} \\
    & I_{C} = C_{J}\frac{d^{2}\phi_{n}}{dt^{2}}, \label{Eq:A2b}\\
    & I_L = I_0 \sin\left(\frac{\phi_n}{\phi_0}\right), \label{Eq:A2c}
\end{align}
\label{Eq:A2}%
\end{subequations}
with $\phi_n$ denoting the node flux associated with the $n$-th node, $I_0$ the critical current of the Josephson junction, and $\phi_0 = \Phi_0/2\pi$ being the reduced flux quantum (with $\Phi_0 = h/2e$). Differentiating Eq.~(\ref{Eq:A2c}) and expanding the result for weak currents ($I_L \ll I_0$), leads to
\begin{align}
    \frac{d\phi_n}{dt} &= \frac{\phi_0}{I_0}\left(1 + \frac{1}{2} \left(\frac{I_L}{I_0}\right)^{2}\right)\frac{dI_L}{dt}. \label{Eq:A3}
\end{align}
Equation~(\ref{Eq:A3}) can be solved for $\phi_n$ as
\begin{align}
    \phi_n &= L_{J} I_L + \frac{L_{J}}{6I_0^{2}}I_L^{3}, \label{Eq:A4}
\end{align}
where we have used $\phi_0 = L_{J} I_0$. Rearranging Eq.~(\ref{Eq:A4}) we can express $I_L$ in terms of $\phi_n$, 
\begin{align}
    I_L &= \frac{\phi_n}{L_{J}} - \frac{\phi_n^{3}}{6I_0^{2}L_{J}^{3}}.
    \label{Eq:A5}
\end{align}
where in the nonlinear term leading order expression for $\phi_n \approx L_{J} I_L$ has been substituted. Putting everything together,
Eq.~(\ref{Eq:A1}) gives
\begin{align}
    C_J\frac{d^{2}\phi_n}{dt^{2}} - C\frac{d^{2}}{dt^{2}}(\phi_{n+1} + \phi_{n-1}-2\phi_{n}) + \frac{\phi_n}{L_{J}} - \frac{\phi_n^{3}}{6I_0^{2}L_{J}^3} = 0.
    \label{Eq:A6}
\end{align}
By taking the continuum limit of the equation above, we obtain Eq.~(3) reported in the main text,
\begin{align}
    C_J\frac{\partial^{2}\phi}{\partial t^{2}} - Ca^{2}\frac{\partial^{4}\phi}{\partial x^{2}\partial t^{2}} + \frac{\phi}{L_{J}} = \frac{\phi^{3}}{6I_0^{2}L_{J}^{3}}.
    \label{Eq:A7}
\end{align}
%
\subsection{Coupled amplitude equations}
%
%
The nonlinear wave equation can be simplified by introducing the plane wave ansatz for $\phi(x,t)$, and then performing harmonic balance at the frequencies of interest (here $\omega_{p}$, $\omega_{s}$ and $\omega_{i}$). To this end, we write $\phi(x,t)$ as
\begin{align}
    \phi(x,t) = \frac{1}{2}\Big(A_p(x)&e^{i(k_p x+\omega_p t)} + A_s(x)e^{i(k_s x +\omega_s t)} \nonumber \\
    &+ A_i(x)e^{i(k_ix+\omega_i t)} + c.c\Big). \label{Eq:A8}
\end{align}
and establish equations of motion for envelope amplitudes $A_{p,s,i}(x)$.
Substituting Eq.~(\ref{Eq:A8}) into Eq.~(\ref{Eq:A7}), making the slowly-varying envelope approximation, i.e. $(\partial^{2}A_m/\partial x^{2} \ll k_m\partial A_m/\partial x)$, 
%
%
and grouping the resonant terms at pump, signal and idler frequencies, Eq.~(\ref{Eq:A7}) reduces to
\begin{subequations}
\begin{align}
    & A_p(x) = A_{p}(0)e^{i\alpha_p x}, \label{Eq:A9a} \\
    & \frac{d A_s(x)}{d x} = i\alpha_s A_s(x) + i\beta_{s}A_{i}^{\star}(x)e^{i\left(\Delta k_L + 2\alpha_p\right)x}, \label{Eq:A9b} \\
    & \frac{d A_i(x)}{d x} = i\alpha_i A_i(x) + i\beta_{i}A_{s}^{\star}(x)e^{i\left(\Delta k_L + 2\alpha_p\right)x}, \label{Eq:A9c}
\end{align}
\label{Eq:A9}%
\end{subequations}
with
\begin{subequations}
\begin{align}
& \alpha_{m} = (2-\delta_{p,m})\frac{\rho\omega_{m}}{v_{g}(\omega_{m})}  \label{eq:A10a}\\
&\beta_{s,i} = \frac{\rho\omega_{s,i}}{v_{g}(\omega_{s,i})}, \label{Eq:A10b}
\end{align}
\label{Eq:A10}%
\end{subequations}
where $\rho$ denotes the nonlinear mixing coefficient determined entirely by pump frequency and amplitude,
\begin{equation}
\rho= \left(\frac{I_p}{I_0}\right)^{2} \frac{Z_{c}^{2}}{16L^{2}\omega_{p}^{2}} = \left(\frac{I_p}{I_0}\right)^{2}\left(\frac{\omega_{0}}{4\omega_{p}}\right)^{2}. \label{Eq:A11} 
\end{equation}
Here $m\in\{p,s,i\}$ and  $Z_{c} $ denotes the characteristic impedance of the line which is chosen to be 50~$\Omega$ to match the typical impedance of microwave circuitry at input and output. Note that we have ignored the dynamical backaction of signal-idler amplitudes on the pump (the so-called ``stiff-pump" approximation, $|A_{s,i}(x)| \ll A_p(x)$) in writing the solution for the pump wave amplitude in Eq.~(\ref{Eq:A13}).
\par
\begin{figure}[b!]
\centering
\includegraphics[width=0.9\columnwidth]{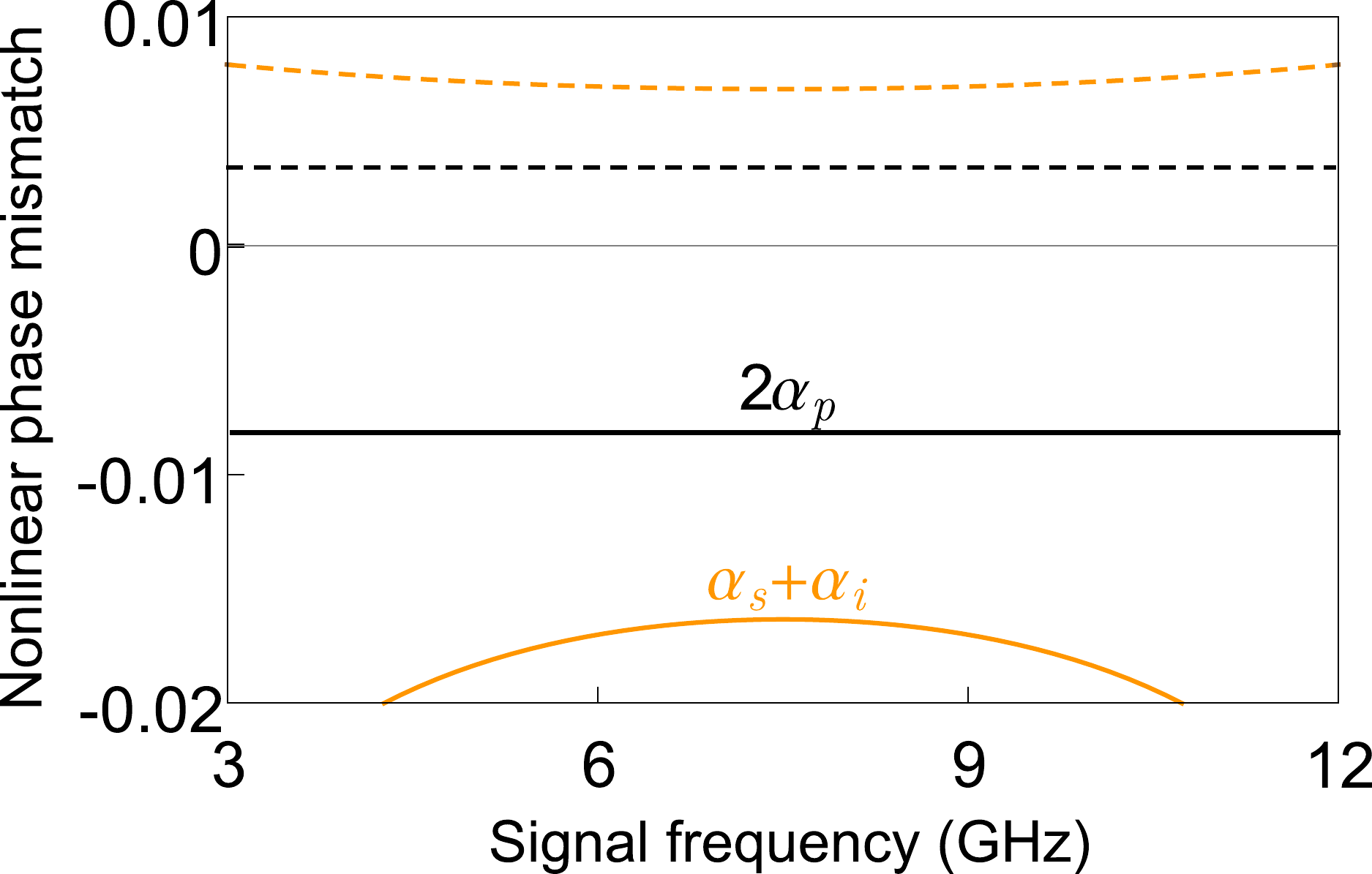}
 \caption{Comparison of nonlinear phase mismatch between left-handed (solid) and right-handed (dashed) J-TWPAs. The black curves show the self-phase modulation coefficient for the pump waves $2\alpha_p$, while the orange curves show the total cross-phase modulation for signal and idler waves $\alpha_s + \alpha_i$. For the left-handed J-TWPA, $2\alpha_p > \alpha_s + \alpha_i$ giving $\Delta k_{NL} > 0$, while the opposite holds for right-handed J-TWPA.}
 \label{Fig:A2}
\end{figure}

Writing the signal and idler amplitudes as ${A_s = a_{s}e^{i\alpha_s x}}$ and ${A_i = a_{i}e^{i\alpha_i x}}$, Eqs.~(\ref{Eq:A10}) simplify to
\begin{subequations}
\begin{align}
    \frac{d a_s}{d x} &= i\beta_{s}a_{i}^{\star}e^{i(\Delta k)x}, \label{Eq:ASeqn}\\
    \frac{d a_i}{d x} &= i\beta_{i}a_{s}^{\star}e^{i(\Delta k)x},
    \label{Eq:AIeqn}
\end{align}
\label{Eq:A12}%
\end{subequations}
where $\Delta k = \Delta k_L + \Delta k_{NL}$,
\begin{subequations}
\begin{align}
    & \Delta k_L = 2k_p - k_s - k_i, \label{Eq:DKL} \\
    & \Delta k_{NL} = 2\alpha_{p} - \alpha_{s} - \alpha_{i}. \label{Eq:DKNL}
\end{align}
\label{Eq:A13}%
\end{subequations}
Figure~\ref{Fig:A2} shows plots of $\alpha_{p,s,i}$ for the left-handed and right-handed J-TWPA respectively with the same circuit parameters as used in Fig.~2 of the main text. Note that for the right-handed case, $2\alpha_p < \alpha_s + \alpha_i$, hence $\Delta k_{NL} < 0$ as per Eq.~(\ref{Eq:DKNL}). However, in the left-handed case, $2\alpha_p > \alpha_s + \alpha_i$ resulting in $\Delta k_{NL} > 0$. 
%
\section{Signal-to-pump backaction and non-degenerate pumping}
\label{App:softpump}
%
%
\subsection{Dual pump}
%
%
Including the signal backaction on the pump necessitates solving the system of equations that includes the spatial variation of pump amplitudes $A_p(x)$ due to signal/idler backaction. For the most general case describing four-wave mixing, these equations are
\begin{subequations}
\begin{align}
    \frac{dA_1(x)}{dx} &= i\beta_1 A_2^{*}(x)A_{3}(x)A_{4}(x)e^{-i\Delta k_{L}x} \nonumber\\
    & \hspace{1cm} + iA_{1}\sum_{m=1}^{4}\alpha_{1m}A_{m}(x)A_{m}^{*}(x), \label{Eq:B1a} \\
    \frac{dA_2(x)}{dx} &= i\beta_2 A_1^{*}(x)A_{3}(x)A_{4}(x)e^{-i\Delta k_{L}x} \nonumber\\
    & \hspace{1cm} + iA_{2}\sum_{m=1}^{4}\alpha_{2m}A_{m}(x)A_{m}^{*}(x), \label{Eq:B1b} \\
    \frac{dA_3(x)}{dx} &= i\beta_3 A_{1}(x)A_{2}(x)A_{4}^{*}(x)e^{i\Delta k_{L}x} \nonumber\\
    & \hspace{1cm} + iA_{3}\sum_{m=1}^{4}\alpha_{3m}A_{m}(x)A_{m}^{*}(x), \label{Eq:B1c} \\
    \frac{dA_4(x)}{dx} &= i\beta_4 A_{1}(x)A_{2}(x)A_{3}^{*}(x)e^{i\Delta k_{L}x} \nonumber\\
    & \hspace{1cm} + iA_{4}\sum_{m=1}^{4}\alpha_{4m}A_{m}(x)A_{m}^{*}(x). \label{Eq:B1d}
\end{align}
\label{Eq:B1}%
\end{subequations}
with the linear phase mismatch modified to
\begin{equation}
   \Delta k_L = k_1 + k_2 - k_3 - k_4.
   \label{Eq:B2}%
\end{equation}
The subscripts $\{1,2\}$ now index the two pump waves, while $\{3,4\}$ index the idler and signal waves respectively. The self- and cross- phase modulation coefficients are given by
\begin{subequations}
\begin{align}
    \alpha_{mn} &= \frac{\omega_m}{16I_{0}^{2}L^{2}v_{g}(\omega_{m})}\left(2-\delta_{mn}\right), \\
    \beta_{m} &= \frac{\omega_{m}}{8I_{0}^{2}L^{2}v_{g}(\omega_{m})},
\end{align}
\label{Eq:B3}%
\end{subequations}
where $m,\,n=\{1,~2,~3,~4\}$. 
\par
Similar to the degenerate pump limit, we can first consider developing the solution for the non-degenerate case under the ``stiff-pump" approximation, i.e. ${|A_{1,2}(x)| \gg |A_{3,4}(x)|}$. Under this assumption, we discard the terms which depend on the signal and idler amplitudes while evaluating the spatial propagation of the two pump waves. This simplifies Eqs.~(\ref{Eq:B3})-(\ref{Eq:B4}) as
\begin{subequations}
\begin{align}
    \frac{dA_{1}(x)}{dx} &= i\alpha_{11}A_{1}(x)\big|A_{1}(x)\big|^{2} + i\alpha_{12}A_{1}(x)\big|A_{2}(x)\big|^{2}, \label{Eq:B4a} \\
    \frac{dA_{2}(x)}{dx} &= i\alpha_{21}A_{2}(x)\big|A_{1}(x)\big|^{2} + i\alpha_{22}A_{2}(x)\big|A_{2}(x)\big|^{2}, \label{Eq:B4b}
\end{align}
\label{Eq:B4}%
\end{subequations}
with the respective solutions given by
\begin{equation}
    A_{1}(x) = A_{1}(0)e^{i\widetilde{\alpha}_{p1}x}, \quad
    A_{2}(x) = A_{2}(0)e^{i\widetilde{\alpha}_{p2}x} 
\label{Eq:B5}%
\end{equation}
where 
\begin{subequations}
\begin{align}
    \widetilde{\alpha}_{p1} &\equiv \frac{2}{3} \sum_{k =1,2}\alpha_{1k}\big|A_{k}(0)\big|^{2}
    = \frac{2}{3}\frac{\omega_{1}}{v_{g}(\omega_{1})}(\rho_{1} + 2\rho_{2}),\\
    \widetilde{\alpha}_{p2} &\equiv \frac{2}{3} \sum_{k =1,2}\alpha_{2k}\big|A_{k}(0)\big|^{2} 
    = \frac{2}{3}\frac{\omega_{2}}{v_{g}(\omega_{2})}(\rho_{2} + 2\rho_{1}).
\end{align}
\label{Eq:B6}%
\end{subequations}
Here, $\rho_{1,2}$ denote the nonlinear mixing coefficients corresponding to each of the two pumps,
\begin{equation}
\rho_{1,2} = \frac{1}{2}\left(\frac{I_{1,2}}{I_{0}}\right)^{2}\left(\frac{\omega_{0}}{4\omega_{1,2}}\right)^{2},
\label{Eq:B7}
\end{equation}
where we have used $A_{j}(0) = I_j Z_{c}/(\sqrt{2} \omega_j)$, $j \in \{1, 2\}$ as the respective pump amplitudes at the input. Note that we have introduced a correction factor of 2/3 while defining $\widetilde{\alpha}_{p1},\,\widetilde{\alpha}_{p2}$. This is required to ensure quantitative agreement with the analysis for the single-pump case presented in Appendix~\ref{App:waveeqn}.2; specifically, in the limiting case of single-pump, ${\omega_{1} = \omega_{2} = \omega_{p}}$ and $A_{1}(x) = A_{2}(x) = A_{p}(x)$, the definition in Eqs.~(\ref{Eq:B6}) enforces
\begin{align}
   \widetilde{\alpha}_{p1} + \widetilde{\alpha}_{p2} = 2\alpha_{p},
   \label{Eq:B8}
\end{align}
with $\alpha_{p}$ being the pump modulation coefficient reported in Eq.~(\ref{eq:A10a}) for the single-pump case. 
\par
Similarly, under the ``stiff-pump'' approximation for double pumps, the signal and idler equations simplify to
\begin{subequations}
\begin{align}
    \frac{dA_3(x)}{dx} &=  i \widetilde{\alpha}_{s} A_{3} + i\widetilde{\beta}_s A_{4}^{\star}(x)e^{i(\Delta k_{L} + \widetilde{\alpha}_{p1} + \widetilde{\alpha}_{p2})x},
    \\
    \frac{dA_4(x)}{dx} &= i \widetilde{\alpha}_{i} A_{4} + i\widetilde{\beta}_i A_{3}^{\star}(x)e^{i(\Delta k_{L} + \widetilde{\alpha}_{p1} + \widetilde{\alpha}_{p2})x},
\end{align}
\label{Eq:B9}%
\end{subequations}
with the pump-induced self-phase modulation of the signal/idler waves described by the coefficients,
\begin{subequations}
\begin{align}
    \widetilde{\alpha}_{s} &= \sum_{k =1,2} \alpha_{3k}\big|A_{k}(0)\big|^{2} = (\rho_{1} + \rho_{2})\frac{2\omega_{3}}{v_{g}(\omega_{3})}, \label{Eq:B10a} \\
    \widetilde{\alpha}_{i} &= \sum_{k =1,2} \alpha_{4k}\big|A_{k}(0)\big|^{2} = (\rho_{1} + \rho_{2})\frac{2\omega_{4}}{v_{g}(\omega_{4})}, \label{Eq:B10b} 
\end{align}
\label{Eq:B10}%
\end{subequations}
and the cross-phase modulation of the signal/idler waves described by the coefficients,
\begin{equation}
    \widetilde{\beta}_{s} = \sqrt{\rho_{1}\rho_{2}}\frac{2\omega_{3}}{v_{g}(\omega_{3})}, \quad \widetilde{\beta}_{i} = \sqrt{\rho_{1}\rho_{2}}\frac{2\omega_{4}}{v_{g}(\omega_{4})}.
    \label{Eq:B11}
\end{equation}
As in the single-pump analysis, we have ignored the self- and cross-phase modulation induced by signal and idler waves under the small signal approximation. Following same steps as sketched in Appendix~\ref{App:waveeqn}, \emph{cis}-gain $G_c (x)$ can be calculated with an effective $\tilde{g}$,
\begin{align}
    \widetilde{g} &= \sqrt{\widetilde{\beta}_{s}\widetilde{\beta}_{i}^{\star} - \left(\frac{\Delta k}{2}\right)^{2}},
    \label{Eq:B12}
\end{align}
where, $\Delta k = \Delta k_{L} + \Delta k_{NL}$ as before, but with the nonlinear wave vector mismatch now defined as, 
\begin{equation}
    \Delta k_{NL} = \widetilde{\alpha}_{p1} + \widetilde{\alpha}_{p2} - \widetilde{\alpha}_{s}  - \widetilde{\alpha}_{i}. \label{Eq:B13}
\end{equation}
\par
The calculation of $G_{c}(x)$ reported for the non-degenerate pumping near $\omega_{\rm ZD}$ in Fig.~4 of the main text makes use of the reduced system of equations in Eq.~(\ref{Eq:B9}).
For the dynamic range calculation presented in Fig.~3 of the main text, we used the full system of equations in Eqs.~(\ref{Eq:B1}), along with the modifications to the pump modulation coefficients as defined in Eq.~(\ref{Eq:B6}).
%
%
\textcolor{\mycolor}{
\subsection{Dynamic range comparison of left- and right-handed J-TWPAs}}
%
%
\begin{figure}[b!]
\centering
\includegraphics[width=0.9\columnwidth]{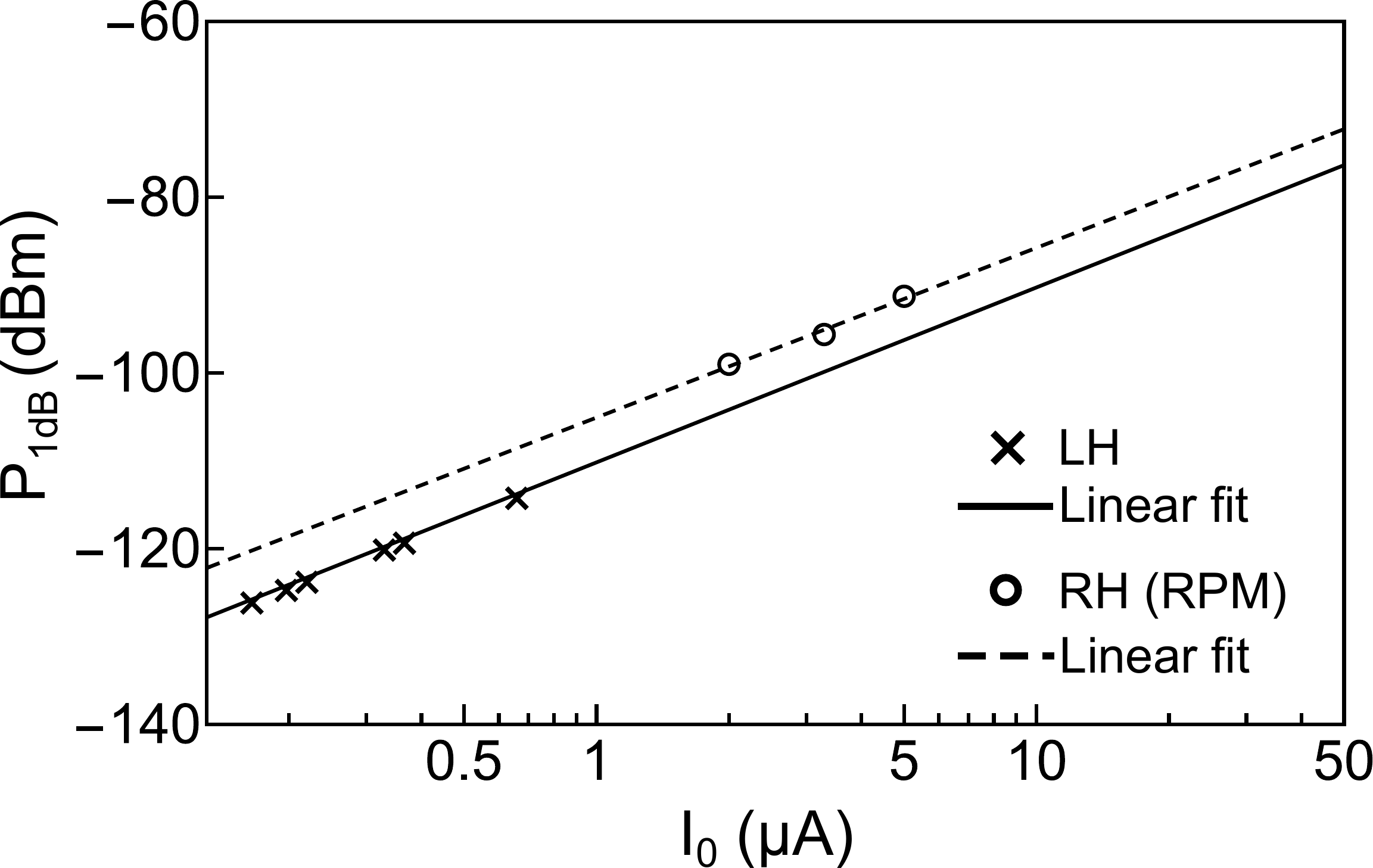}
 \caption{Comparison of $1\text{dB}$ compression power $P_{1\text{dB}}$ in the left-handed and right-handed J-TWPAs as a function of critical current $I_{0}$. To ensure the validity of the continuum approximation, the calculations for the left-handed J-TWPA are performed for smaller values of critical current, leading to an overall smaller 1dB compression power.}
 \label{Fig:S3}%
\end{figure}
\textcolor{\mycolor}{Figure~\ref{Fig:S3} compares the dynamic range for left-handed (crosses) and right-handed (open circles) J-TWPAs as a function of the critical current $I_{0}$ for a peak gain of $20\text{dB}$. The input power corresponding to
$1\text{dB}$ gain compression is approximately $5\text{dB}$ lower in the left-handed device due to the larger junction inductance. 
In general, increasing $I_{0}$ improves the saturation power for both J-TWPA designs, with an increase of around 1.4 dB in $P_{\rm 1\; dB}$ per dB decrease in gain.}

\textcolor{\mycolor}{A relatively standard way of improving the power handling capability, is to use an array of $N$ junctions, with each junction having a critical current value of $NI_{0}$ \cite{Masluk2012}, instead of a single small junction to get the requisite inductance.}
%
\vspace{1cm}
\section{Zero dispersion frequency (ZDF)}
\label{App:ZDF}
%
\begin{figure}[b!]
\centering
\includegraphics[width=0.9\columnwidth]{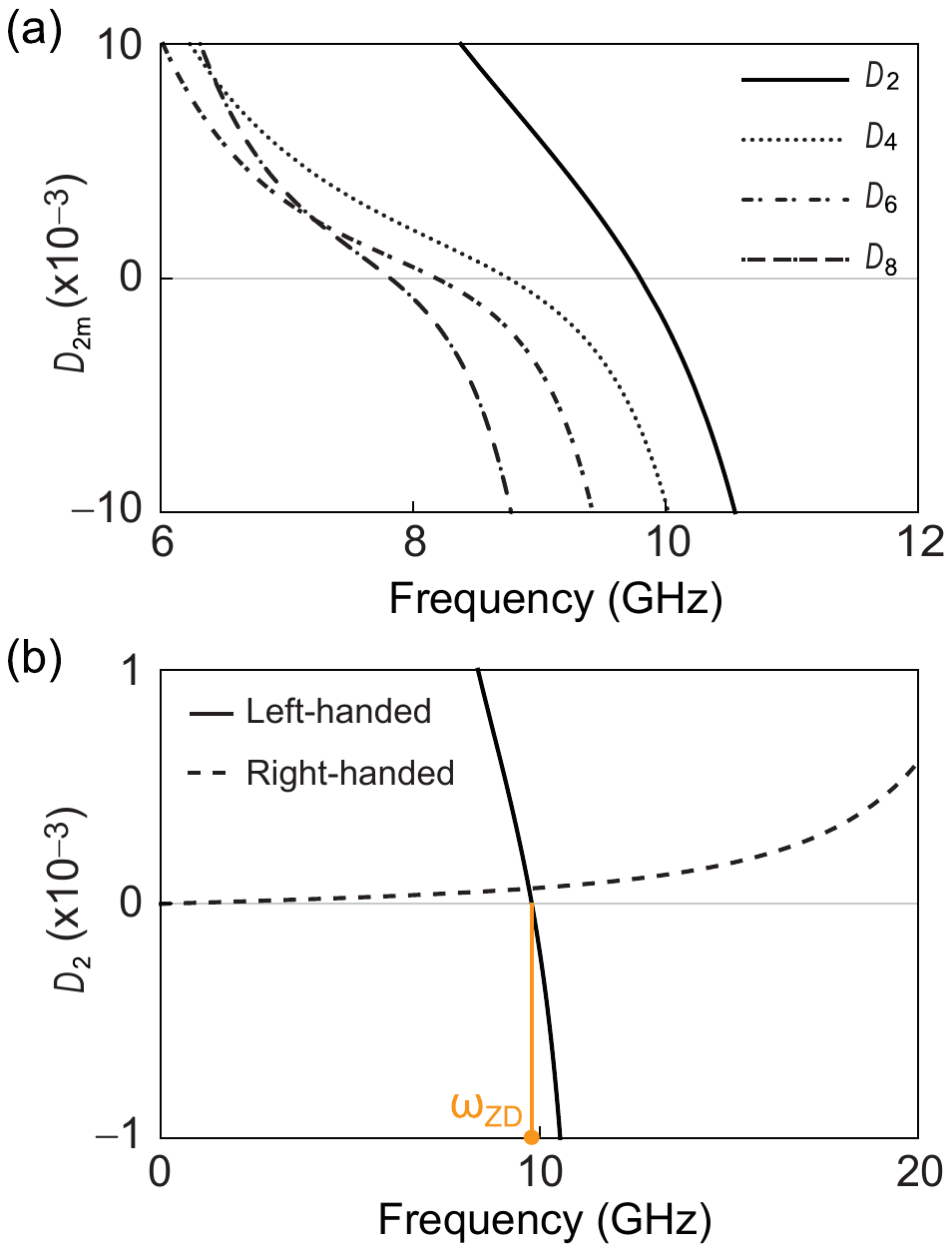}
 \caption{(a) Even-order dispersion parameters $D_{2m}$ as a function of frequency for the left-handed J-TWPA. The circuit parameters used are the same as those used in Fig.~4 of the main text. (b) Comparison of $D_{2}(\omega)$ for left-handed (solid) and right-handed (dashed) JTLs. The circuit parameters for the right-handed J-TWPA are the same as those used in Fig.~2 of the main text. The position of $\omega_{\rm ZD} \approx 9.8\text{GHz}$ is indicated with the orange dot on x-axis.}
 \label{Fig:C1}%
\end{figure}
Expanding $k(\omega)$ around the pump frequency $\omega_p$, 
\begin{equation}
    k(\omega) = k(\omega_{p}) + \left. \sum_{m=1}^{\infty}\frac{D_{m}(\omega)}{m!}\right|_{\omega_{p}}(\omega-\omega_{p})^{m},
    \label{Eq:C1}
\end{equation}
with
\begin{align}
    D_{m} (\omega) &= \frac{d^{m}k(\omega)}{d\omega^{m}}, \label{Eq:C2}
\end{align}
we can write the wave vector at a fixed detuning $\delta$ from the pump, $\omega = \omega_{p}(1 + \delta)$, as
\begin{equation}
    k(\delta) = k(0) + \left. \sum_{m=1}^{\infty}\frac{D_{m}(\delta)}{m!}\right|_{\delta =0}\delta^{m}.
    \label{Eq:C3}
\end{equation}
Using $k_{s} = k(-\delta)$ and $k_{i} = k(\delta)$,
\begin{align}
    \Delta k_{L}(\delta) &= 2k(0) - k(-\delta) - k(\delta) \nonumber\\
     &= -2\sum_{m=1}^{\infty}D_{2m}(\delta)\Big|_{\delta=0}\frac{\delta^{2m}}{(2m)!}, 
     \label{Eq:C4}
\end{align}
where only the even-order dispersion parameters survive \cite{AGRAWAL2019401}. The sign of leading order term $D_{2}(\omega)$, also known as the \emph{group velocity dispersion (GVD)} parameter, determines the sign of $\Delta k_L (\delta)$ at a given detuning.  
\par
Figure~\ref{Fig:C1}(a) shows the frequency dependence of higher even-order dispersion parameters calculated for a left-handed JTL. For frequencies $\omega > \omega_{\rm ZD}$, $D_{2,4,6,8}(\omega) < 0$ leading to $\text{sgn}(\Delta k_L) > 0$, in accordance with Eq.~(\ref{Eq:C4}). Figure~\ref{Fig:C1}(b) compares the GVD parameter $D_{2} (\omega)$ calculated for both the left- and right-handed JTLs. The $\omega_{\rm ZD}$ corresponds to the x-intercept of the the GVD parameter, i.e. $D_{2}(\omega_{\rm ZD})=0$. For a right-handed JTL, ${D_2(\omega) > 0, \, \forall \omega<\omega_{J}}$ preempting the existence of an $\omega_{\rm ZD}$.
%
%
%
\textcolor{\mycolor}{
\section{Non-idealities in a left-handed JTL}
%
\subsection{Effect of higher harmonics}
\label{App:THG}}
%
%
\textcolor{\mycolor}{
To estimate the magnitude of third harmonic generation, we employ the following modified wave ansatz $\phi(x,t)$ in Eq.~(\ref{Eq:A7}),
\begin{subequations}
\begin{align}
    \phi(x,t) &= \frac{1}{2}\Big(A_{p}(x)e^{i(k_{p}x+ \omega_{p}t)}\Big.\nonumber\\
    & \hspace{1cm}+\Big.A_{\text{th}}(x)e^{i(k_{\text{th}}x+\omega_{\text{th}}t)} + c.c\Big),\label{Eq:S31}
\end{align}
\end{subequations}
with $\omega_{\text{TH}} = 3\omega_{p}$. Going to a rotating frame defined as $A_{\text{th}}(x) = a_{\text{th}}e^{i\kappa_{1}x}$, the equations of motion for the amplitudes of the fundamental pump tone $A_{p}(x)$ and the third pump harmonic $a_{\text{th}}(x)$ can be written as,
\begin{subequations}
\begin{align}
    \frac{dA_{p}}{dx} &= i\kappa_{0}A_{p},\label{Eq:S32} \\
    \frac{da_{\text{th}}}{dx} &= -i\kappa_{2}e^{i(\Delta k_{L}+3\kappa_{0}-\kappa_{1})x}.\label{Eq:S33}
\end{align}
\end{subequations}
The solutions to the above equations can be simply written as,
\begin{subequations}
\begin{align}
    A_{p}(x) &= A_{p}(0)e^{i\kappa_{0}x},\label{Eq:S34}\\
    a_{\text{th}}(x) &= \frac{\left(1-e^{i\Delta k x}\right)\kappa_{2}}{\Delta k},\label{Eq:S35}
\end{align}
\end{subequations}
where,
\begin{subequations}
\begin{align}
    \kappa_{0} &= -\frac{\rho k_{p}}{1-\omega_{p}^{2}/\omega_{J}^{2}}, \label{Eq:S36b} \\
    \kappa_{1} &= -\frac{2\rho k_{\text{th}}}{1-\omega_{\text{th}}^{2}/\omega_{J}^{2}},\label{Eq:S36c} \\
    \kappa_{2} &= \frac{\rho k_{\text{th}}}{3\left(1-\omega_{\text{th}}^{2}/\omega_{J}^{2}\right)}A_{p}(0),\label{Eq:S36c}
\end{align}
\end{subequations}
with $A_{p}(0) = I_{p}Z_{c}/\omega_{p}$, $\Delta k = \Delta k_{L} + 3\kappa_{0}-\kappa_{1}$ and $\Delta k_{L} = 3k_{p} - k_{\text{TH}}$. Note that in writing the solutions in Eq.~(\ref{Eq:S35}), we have ignored the effect of backaction from the third harmonic on the fundamental pump tone. Relaxing this `stiff pump' condition leads to enhancement of absolute magnitude of $a_{\text{th}}$, but as far as comparing the relative magnitudes of third harmonic generation between left- and right-handed designs is concerned, a reliable estimate can still be obtained under this simplifying approximation.}
\par
\begin{figure}[t!]
\centering
\includegraphics[width=0.9\columnwidth]{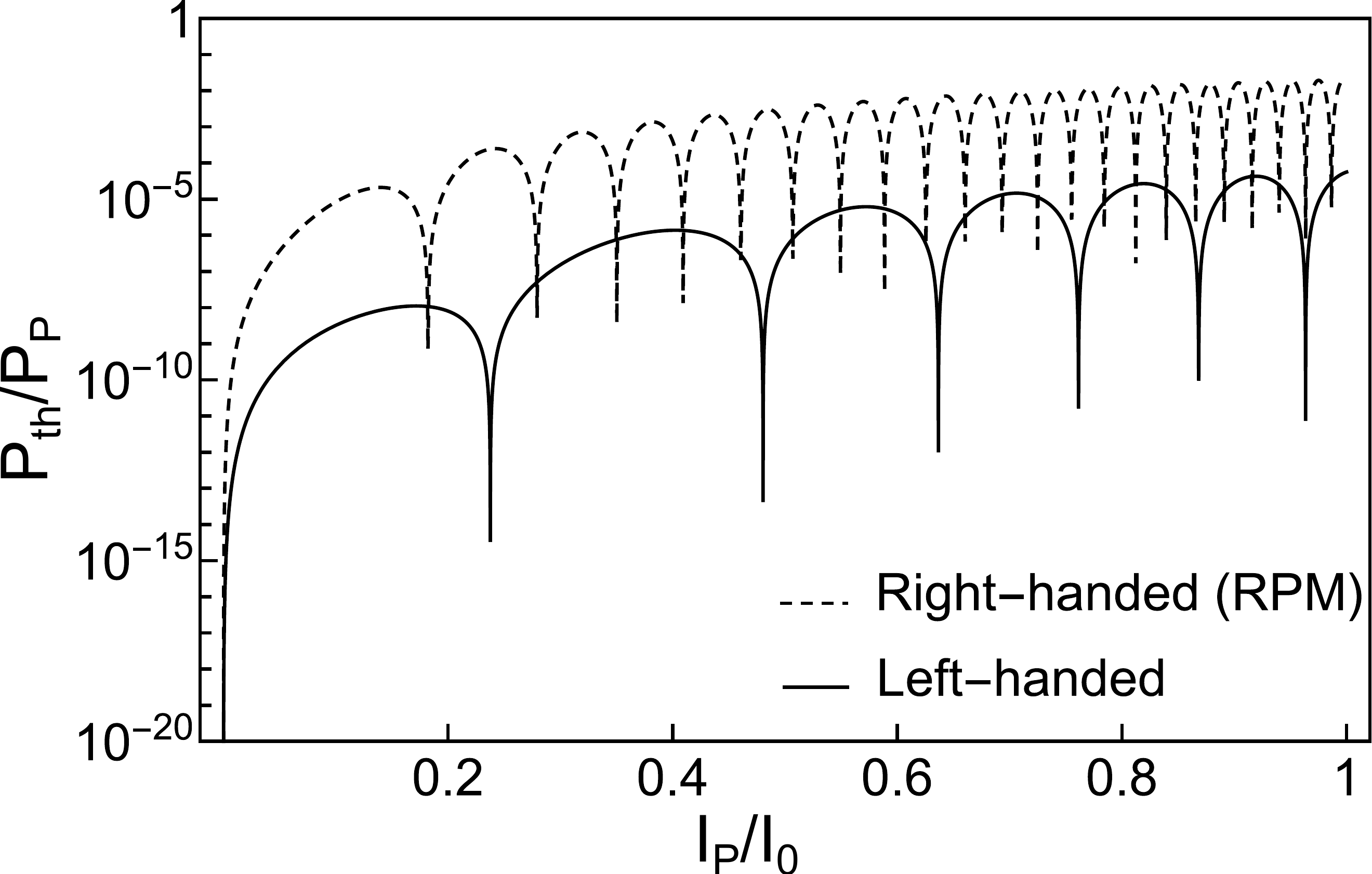}
 \caption{Comparison of normalized third harmonic power $P_{\text{th}}$ for left-handed (solid) and right-handed (dashed) RPM J-TWPAs. Aside from the resonant elements, the circuit parameters for both left- and right- J-TWPAs are the same as those used in Fig.~(2) of the main text. For $I_{p}=0.5I_{0}$, the third harmonic power is approximately three orders of magnitude smaller in the left-handed J-TWPA than the right-handed J-TWPA.}
 \label{Fig:S4}
\end{figure}

\textcolor{\mycolor}{
One can further define the third harmonic power as, $P_{\text{th}} = I_{\text{th}}^{2}Z_{c}/2$
where the current $I_{\text{th}}=\big|a_{\text{th}}(x)\big|\omega_{\text{th}}/Z_{c}$.
Figure~\ref{Fig:S4} compares $P_{\text{th}}$ relative to the fundamental pump power $P_{p}$ for the left- and right- handed J-TWPAs as a function of normalized pump current $I_{p}/I_{0}$. The circuit parameters for the resonant element of the right-handed J-TWPA follows from Ref.~\cite{OBrien2014}. At $I_{p}=0.5I_{0}$, $P_{\text{th}}/P_{p} \approx 10^{-7}$ for the left-handed line whereas $P_{\text{th}}/P_{p} \approx 10^{-4}$ for the right-handed line. Hence the third harmonic power is three orders of magnitude smaller for the left-handed compared to the right-handed JTL.}
\newpage
%
%
%
%
\vspace{1cm}
\textcolor{\mycolor}{
\subsection{Effect of Distributed Loss in a Left-handed J-TWPA}
\label{App:losstangent}
%
%
The primary loss mechanism in J-TWPAs is due to the dielectric loss of capacitance associated with the line. This insertion loss can thus be modeled by adding a fictitious resistor $R$ in parallel to the capacitor $C$ in each unit cell. The modified wave equation from Eq.~(\ref{Eq:A7}) in presence of these resistors can be written as,
\begin{align}
    C_J\frac{\partial^{2}\phi}{\partial t^{2}} - Ca^{2}\frac{\partial^{4}\phi}{\partial x^{2}\partial t^{2}} + \frac{\phi}{L_{J}} - \frac{a^{2}}{R}\frac{\partial^{3}\phi}{\partial t\partial x^{2}} = \frac{\phi^{3}}{6I_0^{2}L_{J}^{3}}, \label{Eq:S38}
\end{align}
where the fourth term on the left-hand side of the equality describes the effect of dielectric loss. By introducing the loss per unit length experienced by the signal and idler modes as  $\Gamma_{s,i}$, Eqs.~(\ref{Eq:A9b})-(\ref{Eq:A9c}) lead to the following set of coupled equations for signal and idler wave amplitudes,
\begin{subequations}
\begin{align}
    &\frac{d A_s(x)}{d x} = -\frac{\Gamma_{s}}{2}A_{s}(x) + i\alpha_s^{\prime} A_s(x) + i\beta_{s}^{\prime}A_{i}^{\star}(x)e^{i\left(\Delta k_L + 2\alpha_p^{\prime}\right)x}, \label{Eq:S39} \\
    & \frac{d A_i(x)}{d x} = -\frac{\Gamma_{i}}{2}A_{i}(x) + i\alpha_i^{\prime} A_i(x) + i\beta_{i}^{\prime}A_{s}^{\star}(x)e^{i\left(\Delta k_L + 2\alpha_p^{\prime}\right)x},\label{Eq:S40}
\end{align}
\end{subequations}
where,
\begin{subequations}
\begin{align}
    \Gamma_{s,i} &= \frac{k_{s,i}\text{tan}\,\delta}{1+\text{tan}^{2}\delta}\left(1-\frac{4\rho}{1 - \omega_{s,i}^{2}/\omega_{J}^{2}}\right),\label{Eq:S41a} \\
    \alpha_{s,i}^{\prime} &= -\frac{k_{s,i}}{2\left(1+\text{tan}^{2}\delta\right)}\left(\text{tan}^{2}\delta + \frac{4\rho}{1 - \omega_{s,i}^{2}/\omega_{J}^{2}}\right),\label{Eq:S41b} \\
    \beta_{s,i}^{\prime} &= -\frac{\rho k_{s,i}}{\left(1 - i\text{tan}\,\delta\right)\left(1 - \omega_{s,i}^{2}/\omega_{J}^{2}\right)}.\label{Eq:S41c}
\end{align}
\end{subequations}
Without loss of generality, we parametrize the value of resistance corresponding to the loss seen at pump frequency, i.e. $\text{tan}\,\delta = 1/\omega_{p}RC$. Adopting a similar method as done for the lossless case, the expression for signal gain in the presence of loss is calculated as
\begin{align}
    G_{c}^{\prime}(x) = \Bigg|\Bigg(\text{cosh}&(g^{\prime}x) + \frac{\Gamma_{i}-\Gamma_{s}}{4g^{\prime}}\text{sinh}(g^{\prime}x)\nonumber \\
     &-\frac{i\Delta k^{\prime}}{2g^{\prime}}\text{sinh}(g^{\prime}x)\Bigg) e^{-(\Gamma_{s}+\Gamma_{i})x/4}\Bigg|^{2}.\label{Eq:S42}
\end{align}
Here, $g^{\prime}$ denotes the modified gain per unit length in the presence of loss,
\begin{align}
    g^{\prime} = \sqrt{\beta_{s}^{\prime}\beta_{i}^{\prime\star}+\left(\frac{\Gamma_{i}-\Gamma_{s}-2i\Delta k^{\prime}}{4}\right)^{2}}, \label{Eq:S43}
\end{align}
with $\Delta k^{\prime}_{NL} = 2\alpha_{p}^{\prime} - \alpha_{s}^{\prime} - \alpha_{i}^{\prime}$ and $\Delta k^{\prime} = \Delta k_{L} + \Delta k_{NL}^{\prime}$. We note that the form of the expression for signal gain in the presence of loss is identical to that obtained for a right-handed J-TWPA in presence of distributed loss \cite{Houde2019}, even though the details of frequency dependence of various parameters entering the expression are obviously different.}
\par
\begin{figure}[t!]
\centering
\includegraphics[width=0.9\columnwidth]{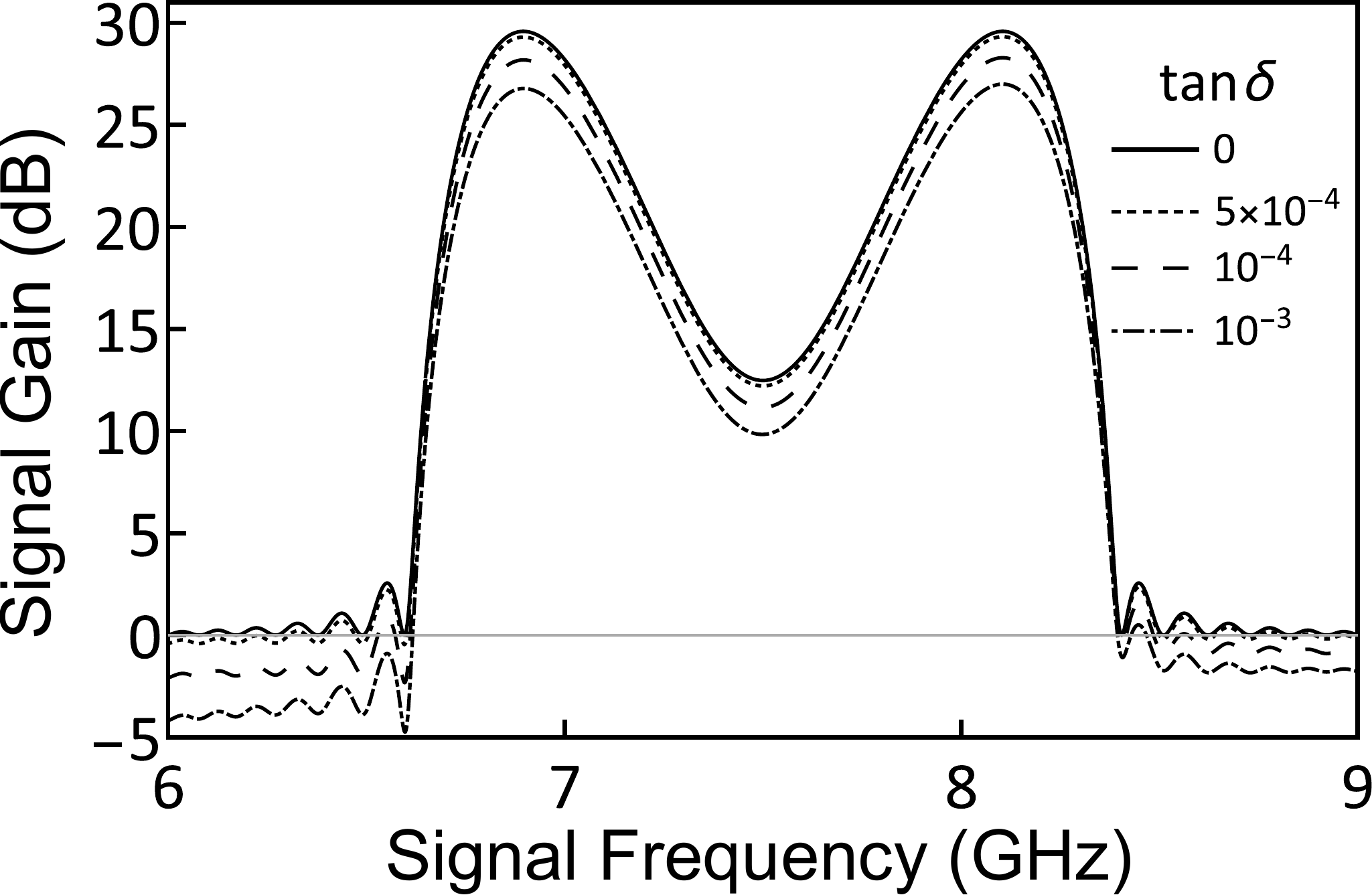}
 \caption{Frequency-dependence of signal gain for different magnitudes of loss in a left-handed JTWPA. The circuit parameters used are similar to those used in Fig.~2 of the main text. For $\text{tan}\,\delta = 10^{-3}$, the peak gain drops by around $3\text{dB}$.}
 \label{Fig:S5}%
\end{figure}

\textcolor{\mycolor}{
As shown in Fig.~(\ref{Fig:S5}), for a typical (and not entirely optimistic!) value for dielectric loss tangent, $\text{tan}\,\delta = 10^{-3}$ \cite{Macklin2015}, the insertion loss for left-handed J-TWPA at $\delta_{max}$ is $\approx 3\text{dB}$. In contrast, for a right-handed JTL of length $\approx1000a$ and $\text{tan}\,\delta = 10^{-3}$, the insertion loss is $\approx1.3\text{dB}$ \cite{Macklin2015}, dominantly contributed by the dielectric of the capacitor forming the RPM element. Given the higher gain per unit length for left-handed JTL, for a fixed peak gain, some of this loss can be compensated by using a shorter left-handed J-TWPA.}
\newpage
%
%
%

%